\documentclass[11pt,a4paper]{article}
\usepackage{jstyle}
\usepackage{tikz}
\usepackage{amsthm,mathrsfs,stmaryrd,bm}
\usepackage[vcentermath]{youngtab}
\usepackage{simplewick}
\usepackage{framed,pifont}

\newcommand{\tr}{{\rm Tr}}
\newcommand{\lstar}{\raisebox{0.2ex}{$\,\st\bigstar\,$}}
\def\a{\alpha}
\def\b{\beta}

\def\d{\delta}
\def\D{\Delta}
\def\e{\epsilon}

\def\l{\lambda}
\def\L{\Lambda}

\def\n{\nu}

\def\r{\rho}
\def\s{\sigma}
\def\S{\Sigma}
\def\t{\tau}

\def\O{\Omega}
\def\o{\omega}

\author[a]{Euihun JOUNG}
\author[b]{\quad Karapet MKRTCHYAN}
\affiliation[a]{AstroParticule et Cosmologie\footnote{UMR 7164
(CNRS, Universit\'e Paris 7, CEA, Observatoire de Paris)}\\
10 rue Alice Domon et L\'eonie Duquet, 75205 Paris Cedex 13, France}
\affiliation[b]{Scuola Normale Superiore and INFN\\
Piazza dei Cavalieri 7, 56126 Pisa, Italy}
\emailAdd{joung@apc.univ-paris7.fr}
\emailAdd{karapet.mkrtchyan@sns.it}

\title{\centering
Notes on higher-spin algebras:\\
{\LARGE  minimal representations and structure constants}}

\abstract{The higher-spin (HS) algebras relevant to Vasiliev's equations in various dimensions
can be interpreted as the symmetries of the \emph{minimal representation}
of the isometry algebra.
After discussing this connection briefly, we generalize this concept
to any classical Lie algebras and consider the corresponding
HS algebras. For $\mathfrak{sp}_{2N}$ and $\mathfrak{so}_{N}$\,,
the minimal representations are unique so we get unique HS algebras.
For $\mathfrak{sl}_{N}$\,, the minimal representation has one-parameter family, so does the corresponding HS algebra. The $\mathfrak{so}_{N}$ HS algebra is what underlies the Vasiliev theory
while the  $\mathfrak{sl}_{2}$ one coincides with the $3D$ HS algebra $hs[\lambda]$\,.
Finally, we derive the explicit expression of the structure constant
of these algebras --- more precisely, their bilinear and trilinear forms.
Several consistency checks are carried out for our results.

}

\begin{document}

\maketitle

\section{Introduction}
\label{sec: intro}

Gauge theories describing massless particles
are naturally endowed with Lie algebras as the background preserving
part of gauge symmetries, namely the global symmetries.
Inversely, one may start with a proper Lie algebra
as a global symmetry and then obtain the field theory by gauging it.
One of the simplest non-trivial examples would be Gravity where
the gauge symmetries --- diffeomorphisms --- give rise to
the isometry algebras (Poincar\'e, AdS or dS) as global symmetries.
The Cartan formulation of Gravity makes use of the inverse construction.
Hence, in the study of massless higher-spin (HS) particles, it is also
of primary importance to investigate the underlying global symmetries ---
HS algebras. Indeed, HS algebra is in the core of Vasiliev's equations \cite{Vasiliev:1990en,Vasiliev:2003ev,Vasiliev:2004qz}
and recent developments of three-dimensional HS theories
--- see the review \cite{Gaberdiel:2012uj} and references therein.

HS algebra is, by definition, the Lie algebra of the global symmetries
underlying a theory involving HS spectrum.
However, in this paper, we shall use a looser definition for the term
\emph{HS algebra} which will be specified later in the text.
 So far, only one HS algebra is known to be fully consistent in each dimensions $D\ge 4$\,.
 The four-dimensional one \cite{Fradkin:1986ka} was considered by Fradkin and Vasiliev in
 the construction of HS cubic interactions \cite{Fradkin:1987ks}, and then used
 for the interacting theory --- Vasiliev's equations.
Extensions to five and seven dimensions have been studied respectively in \cite{Sezgin:2001zs,Vasiliev:2001wa} and \cite{Sezgin:2001ij}.
Finally, the generalization to any dimensions of the algebra together with
 the equations was carried out in \cite{Vasiliev:2003ev,Vasiliev:2004qz}.
These constructions were based on oscillators ---
spinors for four and five dimensions and vectors for any dimensions.
It is important to note that
this HS algebra is also isomorphic to the conformal HS symmetries of the free massless scalar in \mt{D-1} dimensions \cite{Mikhailov:2002bp,Eastwood:2002su}.
Three-dimensional case is special: there is one-parameter family of HS algebras $hs[\lambda]$
\cite{Bergshoeff:1989ns,Bordemann:1989zi,Vasiliev:1989re,Pope:1989sr,
Pope:1990kc,Fradkin:1990ir} corresponding to the
one-parameter family of backgrounds of the $3D$ interacting equations
\cite{Prokushkin:1998vn,Prokushkin:1998bq}.\footnote{Recently, the
 asymptotic symmetries of the $hs[\lambda]\oplus hs[\lambda]$ Chern-Simons theories are identified with the W-algebras $\mathcal W_{\infty}[\lambda]\oplus \mathcal W_{\infty}[\lambda]$ \cite{Henneaux:2010xg,Campoleoni:2010zq, Gaberdiel:2011wb,Campoleoni:2011hg}.}
Besides the explicit construction through oscillators,
the aforementioned HS algebras can be also obtained as
a quotient of universal enveloping algebra (UEA)
of the isometry algebra.\footnote{This view
has been discussed in \cite{Iazeolla:2008ix},
and pursuing the same idea a class of mixed-symmetry HS algebras has been considered in \cite{Boulanger:2011se}.
See also \cite{Fradkin:1989yd,Bekaert:2007mi,Bekaert:2008sa,Bekaert:2011js,Maldacena:2011jn,Boulanger:2013zza,Alba:2013yda,Stanev:2013qra} for other discussions
on HS algebras, and \cite{Eastwood:2006, Gover:2009,
Michel:2011,Bekaert:2013zya}
and
\cite{Vasiliev:2012tv,Gelfond:2013xt}
for generalizations.}

In fact, HS algebra as a coset of UEA has a deep relation
to the theory of \emph{minimal representations}.
In 1974, Joseph \cite{Joseph:1974} raised the question that what is the minimum number of Heisenberg pairs which are needed to represent a Lie algebra.
To attack this question, he coined
the notion of \emph{Joseph ideal} which corresponds to the kernel of the aforementioned representation --- minimal representation.
The coset of UEA by Joseph ideal is an infinite-dimensional
algebra including the original Lie algebra as subalgebra.
Interestingly, HS algebras mentioned in the above paragraph all fall into
these coset Lie algebras.
Much progress has been made on the subject of  minimal representations,
see for example \cite{Joseph:1976,Garfinkle:1982,Levasseur:1988,Binegar:1991,Braverman:1998,Li:2000,Gan:2004,Eastwood:2007,Fronsdal:2009,Todorov:2010md,Kobayashi:2013ao} and references therein.
In physics literature, the minimal representations of isometry algebras
are explored to a large extent by Gunaydin and collaborators \cite{Gunaydin:1989um,Fernando:2009fq,Fernando:2010dp,Govil:2013uta,Govil:2014uwa}.

In this paper, we first make a brief survey of the construction of
HS algebras from the point of view of minimal representations.
This construction is technically not much new compared to what was known in the HS literature, and we just attempt to make a link between two rather disconnected
literatures. As we mentioned we shall use the term \emph{HS algebra}
loosely, as the symmetry algebra of the minimal representation of a finite-dimensional Lie algebra.
In this article, we shall focus on the classical Lie algebras, $\mathfrak{sl}_{N},
\mathfrak{so}_{N}$ and $\mathfrak{sp}_{2N}$\,.
The original part of the present paper, in a strict sense,
is the presentation of the explicit structure constants of the HS algebras.
In order to present them, let us first introduce the relevant notations.
Let $T_{\bm a}$ denote the generators of a Lie algebra, say $\mathfrak g$\,.
Then, the corresponding HS algebra, denoted by $hs(\mathfrak g)$\,, can be given through an arbitrary function of
the element $A^{\bm a}$ in $\mathfrak g^{*}$\,, the dual space of $\mathfrak{g}$\,:
\be
	T(A)=\sum_{n=0}^{\infty}\,\frac1{n!}\,T_{\bm a_{1}\cdots\,\bm a_{n}}\,A^{\bm a_{1}}\cdots\,A^{\bm a_{n}}\,.
\ee
The coefficients $T_{\bm a_{1}\cdots\,\bm a_{n}}$ correspond to the generators
of $hs(\mathfrak{g})$.
To be more precise, $A^{\bm a}$ are not arbitrary elements of
$\mathfrak g^{*}$\,,
but belong to its minimal orbit whose precise conditions will be presented later.
Due to such conditions, the generators  $T_{\bm a_{1}\cdots\,\bm a_{n}}$
have a less number of independent components,
and they can be obtained from the
Taylor expansion of $T(A)$\,.
Hence,
the algebraic structure of $hs(\mathfrak{g})$
can be studied in terms of $T(A)$\,:
\begin{itemize}
\item
 We first consider the bilinear form
 $\mathcal B(A)=\tr[\,T(A_{1})\lstar T(A_{2})\,]$
 where $\lstar$ and $\tr[\,\cdot\,]$ denote respectively the associative product
 and the trace operation of $hs(\mathfrak{g})$
whose definitions will be provided later.
This bilinear form serves as a HS generalization of the Killing form.
\item Then, we
move to the trilinear form $\mathcal C(A)=\tr[\,T(A_{1})\lstar T(A_{2})\lstar T(A_{3})\,]$
from which the structure constant of the algebra, consisting of totally symmetric
and antisymmetric parts,  can be obtained.
\end{itemize}
For the HS algebras associated with classical Lie algebras,
we find
\begin{framed}
\vspace{-10pt}
\paragraph{\underline{$\mathfrak{sp}_{2N}$ series}}
\be
	\mathcal B(U)=\frac{1}{\sqrt{1+\frac{\la U_{1}\,U_{2}\ra}4}}\,,
	\qquad
	\mathcal C(U)=\frac{1}{\sqrt{1+\frac{
	\la U_{1}\,U_{2}\ra+\la U_{2}\,U_{3}\ra+\la U_{3}\,U_{1}\ra
	+\la U_{1}\,U_{2}\,U_{3}\ra}4}}\,.
	\label{BCU}
\ee
\paragraph{\underline{$\mathfrak{sl}_{N}$ series}}
\ba
	\mathcal B(V) \eq {}_3F_2\Big(\tfrac{N}2\,(1+\l)\,,\,
	\tfrac{N}2\,(1-\l)\,,\,1\,;\,
\tfrac{N}2\,,\,\tfrac{N+1}2\,;\,-\tfrac14\,\la V_{1}\,V_{2}\ra\Big)\,,
	\\
	\label{C(V)}
	\mathcal C(V) \eq \sum_{k=0}^{\infty}\,\sum_{\ell=0}^{k}\,
(-1)^{k}\,\binom{k}{\ell}\,
\frac{\big(\tfrac{N(1+\l)}2\big)_{2\,k-\ell}
\big(\tfrac{N(1-\l)}2\big)_{k+\ell}}
{(N)_{3k}}\times\nn
&& \qquad \times\,\big[\la V_{1}\,V_{2}\ra+\la V_{2}\,V_{3}\ra+\la V_{3}\,V_{1}\ra+\la V_{1}\,V_{2}\,V_{3}\ra\big]^{k-\ell}\nn
&& \qquad \times\,
\big[\la V_{1}\,V_{2}\ra+\la V_{2}\,V_{3}\ra+\la V_{3}\,V_{1}\ra-\la V_{3}\,V_{2}\,V_{1}\ra\big]^{\ell}\,.
\ea
\paragraph{\underline{$\mathfrak{so}_{N}$ series}}
\ba
	 \mathcal B(W)\eq {}_{2}F_{1}\Big(
	2\,,\,\tfrac{N-4}{2}\,;\,\tfrac{N-1}{2}\,;\,-\tfrac18\,\la W_{1}\,W_{2} \ra \Big)\,,\\
	\mathcal C(W) \eq \sum_{m=0}^{\infty}\sum_{n=0}^{\infty}
	\frac{(-1)^{m}}{m!}\,\frac1{n!}\,
	\frac{\left(\frac{N-4}2\right)_{m+2n}
	(2)_{m+2n}}{\left(\frac{N-1}2\right)_{m+3n}\,8^{m+3n}}\times\nn
	&&\qquad\times\,
	\big[\la W_{1}\,W_{2}\ra+\la W_{2}\,W_{3}\ra
	+\la W_{3}\,W_{1}\ra+\la W_{1}\,W_{2}\,W_{3}\ra \big]^{m} \nn
	&&\qquad \times\, \big[
	\la W_{1}\,W_{2}\ra
	\la W_{2}\,W_{3}\ra \la W_{3}\,W_{1}\ra
	+2\,\la W_{1}\,W_{2}\,W_{3}\ra^{2}\big]^{n}\,.
	\label{C(W)}
\ea
\end{framed}
\noindent Here, $U_{i}$, $V_{i}$ and $W_{i}$ are respectively
elements of ${\mathfrak{sp}_{2N}}^{\!*}$, ${\mathfrak{sl}_{N}}^{\!*}$
 and ${\mathfrak{so}_{N}}^{\!*}$\,, and
 $\la\,\cdot\,\ra$ is the matrix trace.
Various coincidence cases
$\mathfrak{sp}_{2}\simeq \mathfrak{sl}_{2}$,
$\mathfrak{sp}_{4}\simeq \mathfrak{so}_{5}$ and
$\mathfrak{sl}_{4}\simeq \mathfrak{so}_{6}$ can be explicitly checked
from the above formulas.
Only $\mathfrak{sl}_{N}$ series admits an one-parameter family
of HS algebras\footnote{To our best knowledge, this one-parameter family of algebras was discussed first in \cite{Fradkin:1989yd}.}, and
the appearance of ideals for particular values of $\lambda$  is
manifest
from the expression of the bilinear form.
It is worth to notice also that for $\mathfrak{sl}_{2}$, we recover
the $3D$ algebra $hs[\lambda]$ with a  particularly simple form of structure constant, since the trilinear form \eqref{C(V)} can be considerably simplified.

\bigskip

The organization of the paper is as follows:
\begin{itemize}
\item
In Section \ref{sec: gen}, we review some generalities of HS algebras.
First we show how they appear from a field theory of massless HS particles,
and then present their relation to mathematical objects such as
coadjoint orbits and minimal representations.
We provide the definition of
HS algebras
and their realizations in terms of oscillators.

\item
In Section \ref{sec: bt}, we derive the explicit expression of structure constant --- the invariant bilinear and trilinear forms. For that,
we introduce a trace
for an element of HS algebra, defined as the identity piece of the element.
We make explicit the latter definition showing
that such trace can be given in fact through the $\mathfrak{gl}_{1}$ and $\mathfrak{sp}_{2}$ projectors previously introduced
in \cite{Vasiliev:2001wa,Alkalaev:2002rq} and \cite{Vasiliev:2004cm}, respectively for $\mathfrak{sl}_{4}$ and $\mathfrak{so}_{D+1}$\,.
With such trace formulas, we explicitly evaluate the bilinear and trilinear forms ending up with the results (\ref{BCU}--\ref{C(W)}).
\item
In Section \ref{sec: more sl}, we discuss more about the HS algebras associated
with $\mathfrak{sl}_{N}$.
First, we discuss the formation of ideals and
associated finite-dimensional coset algebras,
which arise for particular values of $\lambda$\,.
Then, we provide another description of these HS algebras, based on
reduced number of oscillators which are free from equivalence relations.
At the end of this section, we discuss in more details  the $\mathfrak{sl}_{2}$ case, that is $hs[\lambda]$\,.
Besides providing a relatively simple expression for the $\lstar$ product,
we make a concrete link between the description used in this paper and
that of the deformed oscillators.

\item
Finally, in Section \ref{sec: outlook}, we discuss some outlooks
of the present work,
while Appendix \ref{sec: projector} includes some technical details.

\end{itemize}

\section{HS algebras and Minimal representations}
\label{sec: gen}

In order to keep the current paper as self-complete as possible,
we review
the definition and the construction of HS algebras
collecting knowledge
from the physics and mathematics literature.
Our focus is on providing the precise definition
and role of HS algebras in physics and introducing the notion of minimal representations.

\subsection{HS algebras as Global HS symmetries}
\label{sec: killing}

A HS algebra is the global-symmetry counterpart of HS gauge symmetry.
The latter depends on the description one chooses
--- frame-like, metric-like, etc. --- whereas the global symmetry does not depend on such a choice.
In the following, we briefly introduce HS algebra
as the global symmetry of HS gauge fields in the metric-like description
where the field content is a (infinite) set of symmetric tensor fields
$\varphi_{\mu_{1}\cdots \mu_{s}}$ (with a certain multiplicity for a given spin field).
The gauge transformation has a form,
\be
	\delta_{\varepsilon}\,\varphi_{\mu_{1}\cdots\mu_{s}}
	=\bar\nabla_{(\mu_{1}}\,\varepsilon_{\mu_{2}\cdots \mu_{s})}
	+t_{\mu_{1}\cdots\mu_{s}}(\varphi,\varepsilon)
	+\mathcal{O}(\varphi^{2})\,,
	\label{gauge tr}
\ee
where $\bar \nabla$ is the (A)dS covariant derivative
and $t_{\mu_{1}\cdots \mu_{s}}(\varphi,\varepsilon)$
is the \emph{interaction-part} of transformation which is bilinear in
gauge fields and parameters, denoted by $\varphi$ and $\varepsilon$
respectively.\footnote{To be more precise,
in Fronsdal's formulation of HS fields \cite{Fronsdal:1978rb,Fronsdal:1978vb}, the gauge fields and parameters are subjected
to trace conditions:
\mt{\bar g^{\mu_{1}\mu_{2}}\,\bar g^{\mu_{3}\mu_{4}}\,
	 \varphi_{\mu_{1}\cdots \mu_{s}}=\mathcal O(\varphi^{2})}
and
\mt{\bar g^{\mu\nu}\,\varepsilon_{\mu_{1}\cdots \mu_{s-1}}=\mathcal{O}(\varphi)\,,} where $\bar g^{\mu\nu}$ is the (A)dS inverse metric.}
To restrict ourselves to the global part of such symmetries,
we impose the Killing equations:
\be
	0=\big[\,\delta_{\varepsilon}\,\varphi_{\mu_{1}\cdots\mu_{s}}
	\,\big]_{\varphi=0}
	=
	\bar\nabla_{(\mu_{1}}\,\varepsilon_{\mu_{2}\cdots \mu_{s})}\,.
	\label{Killing}
\ee
The space of the solutions $\bar \varepsilon_{\mu_{1}\cdots \mu_{r}}$
--- Killing tensors ---
defines the HS algebra as a vector space (see \cite{Bekaert:2005ka,Bekaert:2006us} for related works).
For more details,
it is convenient to reiterate the discussion
using auxiliary variables in the ambient-space formulation.
The Killing equation \eqref{Killing} is then given by
\be
	U\cdot \partial_{X}\,E(X,U)=0 \qquad
	[\,X,U\in \mathbb R^{D+1}\,]\,,
\ee
where the ambient-space gauge parameter $E$ is related to
the intrinsic one by
\be
	E(X,U)=\sum_{r=0}^{\infty} \frac{R^{\,r}}{r!}\,
	\bar e_{a_{1}}^{\ \mu_{1}}(x)\,U^{a_{1}}\,\cdots\,
	\bar e_{a_{r}}^{\ \mu_{r}}(x)\,U^{a_{r}}\,\varepsilon_{\mu_{1}\cdots\mu_{r}}(x)\,,
	\label{rel amb}
\ee
with $R$ and $x$ being the radial and (A)dS-intrinsic coordinate
of the ambient space $\mathbb R^{D+1}$\,,
and  $\bar e^{\ \mu}_{a}$ the (A)dS background vielbein.
The relation \eqref{rel amb} is equivalent to imposing
the tangentiality and homogeneity conditions on $E$ as
\be
	X\cdot \partial_{U}\,E(X,U)=0=
	(X\cdot\partial_{X}-U\cdot\partial_{U})\,E(X,U)\,.
\ee
In this ambient-space description, the solution $\bar E$ of the Killing equation
reads simply
\be
	\bar E(X,U)=\sum_{r=0}^{\infty}
	\frac1{2^{r}\,(r!)^{2}}\,\bar E_{a_{1}b_{1},\,\ldots\,,a_{r}b_{r}}\,
	X^{[a_{1}}\,U^{b_{1}]}\,\cdots\,
	X^{[a_{r}}\,U^{b_{r}]}\,,
\ee
and from the tracelessness of the gauge parameter, one
can also conclude that the Killing tensors are completely traceless:
\be
	\partial_{U}^{\,2}\,\bar E(X,U)=0\,,\qquad
	\partial_{X}^{\,2}\,\bar E(X,U)=0\,,\qquad
	\partial_{U}\!\cdot \partial_{X}\,\bar E(X,U)=0\,.
	\label{traceless}
\ee
The generators
of HS algebra are the duals of
$\bar E_{a_{1}b_{1},\,\ldots\,,a_{r}b_{r}}$ and  given by
 \ba
	(M^{a_{1}b_{1},\,\ldots\,,a_{r}b_{r}})(X,U)
	\eq X^{[a_{1}}\,U^{b_{1}]}\,\cdots\,
	X^{[a_{r}}\,U^{b_{r}]}
	+X\cdot U\,S_{1}^{a_{1}b_{1},\,\ldots\,,a_{r}b_{r}}\nn
	&&\hspace{20pt}
	+\,X^{2}\,S_{2}^{a_{1}b_{1},\,\ldots\,,a_{r}b_{r}}
	+ U^{2}\,S_{3}^{a_{1}b_{1},\,\ldots\,,a_{r}b_{r}}\,,
\ea
with arbitrary tensors $S_{i}$ due to the tracelessness
of $\bar E$\,: using such a freedom, one can choose traceless
$M^{a_{1}b_{1},\,\ldots\,,a_{r}b_{r}}$\,.
So far, we have not used any information coming from interactions
but just the field content,
and
we have determined only the vector-space structure of
HS algebra: the basis elements have the symmetry of the rectangular two-row \mt{O(D+1)} diagrams,
\be
	M^{a_{1}b_{1},\,\ldots\,,a_{r}b_{r}}\sim
	{\footnotesize\yng(7,7)}_{\,\circ}\ ,
	\label{M yd}
\ee
that satisfy
\ba
	&M^{\ldots\,,a_{i}b_{i},\,\ldots\,, a_{j}b_{j},\,\ldots}
	=M^{\ldots\,,a_{j}b_{j},\,\ldots\,, a_{i}b_{i},\,\ldots}\,,
	\qquad
	M^{(a_{1}b_{1}),\,\ldots}=0
	=M^{[a_{1}b_{1},a_{2}]b_{2},\,\ldots}\,,\nn
	&\eta_{a_{1}a_{2}}\,M^{a_{1}b_{1},a_{2}b_{2},\,\ldots}=0\,.
\ea
The Lie-algebra bracket $[\![\ ,\, ]\!]$ of HS algebra is inherited from that of
the gauge algebra as
\be
 \delta^{\sst (0)}_{[\![ \varepsilon_{1},\varepsilon_{2}]\!]}=
	\delta^{\sst (0)}_{\varepsilon_{1}}\,\delta^{\sst (1)}_{\varepsilon_{2}}
	-\delta^{\sst (0)}_{\varepsilon_{2}}\,\delta^{\sst (1)}_{\varepsilon_{1}}\,,
\ee
where $\delta^{\sst (0)}_{\varepsilon}$ and $\delta^{\sst (1)}_{\varepsilon}$ are respectively
the first and second terms of the gauge transformation \eqref{gauge tr}.
Hence, the bracket of HS algebra is entirely specified by the first-order interacting
terms $t_{\mu_{1}\cdots \mu_{s}}$ of the gauge transformations,
and they are in turn fixed by the cubic interaction terms of the Lagrangian
--- see \cite{Joung:2013nma} for a recent related discussion.
It is important to note that the global symmetries close at the level of $\delta^{\sst (1)}$\,:
\be
	\delta^{\sst (1)}_{\bar \varepsilon_{1}}\,\delta^{\sst (1)}_{\bar \varepsilon_{2}}
	-\delta^{\sst (1)}_{\bar \varepsilon_{2}}\,\delta^{\sst (1)}_{\bar \varepsilon_{1}}
	=\delta^{\sst (1)}_{[\![ \bar \varepsilon_{1},\bar \varepsilon_{2}]\!]}
	+({\rm trivial\ part})\,,
\ee
so that $\delta^{\sst (1)}_{\bar \varepsilon}$
provides the representation of HS algebra carried by the field content.
Here, (trivial part) means the transformations, either of the form
of free gauge symmetry or proportional to the free equations of motion.
Moreover, this representation leaves the quadratic action $S^{\sst (2)}$ invariant:
$\delta^{\sst (1)}_{\bar \varepsilon}\,S^{\sst (2)}[\varphi]=0$\,,
so is endowed with an invariant scalar product,
which is positive definite if the free action $S^{\sst (2)}$ is unitary.
Hence, for a unitary HS theory, the representation of
HS algebra given by $\delta^{\sst (1)}_{\bar \varepsilon}$ is also unitary.
This condition, known as \emph{admissibility condition} \cite{Konstein:1988yg,Konstein:1989ij},
turns out to be quite tight for
a quest of candidate HS algebras.

Another condition on HS algebra is the requirement that its spin-two part
reproduce Gravity.
This condition fixes certain brackets of HS algebra.
First, the spin-two part gives the isometry algebra $\mathfrak{so}_{D+1}$\,:
\be
	[\![\,M_{ab}\,,M_{cd}\,]\!]=
	2\,(\eta_{a[c}\,M_{d]b}-\eta_{b[c}\,M_{d]a})\,,
	\label{so CR}
\ee
and other HS generators are subject to covariant transformation under isometry:
\be
	[\![\,M_{a_{1}b_{1},\ldots,a_{r}b_{r}}\,,M_{cd}\,]\!]=
	2\,\sum_{k=1}^{r}
	\eta_{a_{k}[c}\,M_{\ldots, d]b_{k},\ldots }
	-\eta_{b_{k}[c}\,M_{\ldots, d]a_{k},\ldots }\,.
	\label{HS commut}
\ee
All in all, apart from the admissibility condition,
any Lie algebra generated by
Killing tensors \eqref{M yd} which transform covariantly under
the isometry algebra $\mathfrak{so}_{D+1}$
is a candidate HS algebra.

\subsection{UEA construction of HS algebras}
\label{sec: UEA}

As discussed in \cite{Iazeolla:2008ix,Boulanger:2011se}, the generators of HS algebra subject to
$\mathfrak{so}_{D+1}$-covariant transformation
can be constructed from the universal enveloping algebra (UEA) of
$\mathfrak{so}_{D+1}$.
In the following discussion, let us focus on the  $D\ge 4$ cases.
The UEA is defined as the quotient of the tensor algebra of $\mathfrak{so}_{D+1}$
with the two-sided ideal generated by
\be
	I_{abcd}=M_{ab}\otimes M_{cd}-M_{cd}\otimes M_{ab}
	-[\![\,M_{ab}\,,\,M_{cd}\,]\!]\,.
\ee
Hence, the class representatives can be taken as $GL(D+1)$ tensors,
\be
	M_{a_{1}b_{1}}\odot\cdots\odot M_{a_{n}b_{n}}\nn
	:=
	\frac1{n!}\sum_{\s \in S_{n}} M_{a_{\s(1)}b_{\s(1)}}\otimes\cdots\otimes M_{a_{\s(n)}b_{\s(n)}}\,,
\ee
which are generically reducible under index-permutation symmetries.
When decomposed into irreducible components,
they contain Killing tensors
 $M_{a_{1}b_{1},\,\ldots\,,a_{n}b_{n}}$
 as well as other elements.
At the quadratic level, the $GL(D+1)$ decomposition gives
\be\label{tensorstructures}
	M^{(a_{1}}{}_{(b_{1}}\odot M^{a_{2})}{}_{b_{2})}
	\sim {\tiny\yng(2,2)}\ ,
	\qquad
	M_{[a_{1}b_{1}}\odot M_{a_{2}b_{2}]}\sim
	{\tiny\yng(1,1,1,1)}\ .
\ee
The traceless part of $M^{(a_{1}}{}_{(b_{1}}\odot M^{a_{2})}{}_{b_{2})}$
gives a Killing tensor
but its trace part and $M_{[a_{1}b_{1}}\odot M_{a_{2}b_{2}]}$
are not Killing tensors.
However, they generate an ideal called
\emph{Joseph ideal}
which we shall discuss more extensively in the next subsection.
When the UEA is quotiented by this ideal,
the coset is spanned only by Killing tensors
so satisfies the condition of HS algebra. To summarize, the following two classes of the elements in $\mathfrak{so}_{D+1}\odot\mathfrak{so}_{D+1}$\,:
\be
	J_{ab}:= M_{(a}{}^{c}{}\odot M_{b)c}-\frac{\eta_{ab}}{D+1}\,
	M^{cd}\odot M_{cd}\sim {\tiny\yng(2)}_{\,\circ}\,,
	\qquad
	J_{abcd} := M_{[ab}\odot M_{cd]}
	\sim {\tiny\yng(1,1,1,1)}\ ,
\ee
generate the Joseph ideal which contains all the
non-Killing-tensor elements.
This construction fixes the values of all $\mathfrak{so}_{D+1}$ Casimir operators, and in particular
the quadratic one is given by
\be
	C_{2}:=\frac12\,M^{ab}\odot M_{ba}=-\frac{(D+1)(D-3)}4\,.
\ee
In  \cite{Boulanger:2011se}, the authors considered another ideal generated only by $J_{ab}$
--- but not by $J_{abcd}$\,.
In such a case,
the quadratic Casimir remains arbitrary while the other Casimirs
are determined as functions of the former.
Hence, one can further take the quotient with
$C_{2}-\nu$\,.
Then, the resulting coset algebra contains
more generators than the original case, and additional generators
correspond to the Killing tensors of certain types of mixed-symmetry fields.
This algebra has one-parameter family with label $\nu$\,,
and it is denoted by $hs(\nu)$\,.
In \mt{D=5} case, $hs(\nu)$ can be decomposed into two parts
which are isomorphic to each other:
each part can be independently obtained
from the ideal generated by the elements $J_{ab}$ and $J_{abcd}^{\pm\lambda}$\,,
where the latter element is
given by
\be
 	J^{\lambda}_{abcd}
	:=M_{[ab}\odot M_{cd]}
	-i\,\frac{\lambda}{6}\,\epsilon_{abcdef}
	\,M^{ef}\,.
	\label{lambda J}
\ee
The generators of the resulting coset algebra
are all given by Killing tensors like the $\lambda=0$ case, and
the quadratic Casimir is given by
\be
	C_{2}=\nu=3\,(\lambda^2-1)\,.
	\label{nu lambda}
\ee
Moreover, one can show that the algebra admits  further ideals
for any half-integer values of $\lambda$ with $|\lambda|\ge1$\,,
and quotienting those ideals results in finite-dimensional algebras.
These finite-dimensional HS algebras have been
considered recently in \cite{Manvelyan:2013oua}
--- see Section \ref{sec: trun} for more details.

To conclude this section, let us also mention that the aforementioned HS algebra can be
obtained by quotienting directly the tensor algebra of $\frak{so}_{D+1}$
with the ideal generated by the following elements:
\ba
    I^{\lambda}_{abcd}\eq
	M_{a[b}\otimes M_{cd]}
	-\eta_{a[b}\,M_{cd]} - i\,\frac{\lambda}{6}\,\e_{abcdef}\,M^{ef}\,,
	\label{I 4}\\
	I^{\lambda}_{ab}\eq M_{c(a}\otimes M_{b)}{}^{c}+
	\frac{D-3}{2}(1-\l^{2})\,\eta_{ab}\label{I lambda}\,,
\ea
where the $\lambda\neq0$ cases are only for $D=5$\,.
From the above, it becomes more clear that Killing tensors
can be taken as class representatives.
The existence of one-parameter family for \mt{D=5}
is actually due to the fact that $\mathfrak{so}_{6}$
is isomorphic to $\mathfrak{sl}_{4}$\,.
There, the elements \eqref{I 4} and \eqref{I lambda}
can be combined into
\be\label{qid}
	{I^{\lambda}\,}^{ac}_{bd}=
	L^{[a}{}_{b}\,\otimes\,L^{c]}{}_{d}
	+\delta^{[a}_{(b}\,L^{c]}{}_{d)} + \lambda\,\delta^{[a}_{[b}\, L^{c]}{}_{d]} + \tfrac14\,(\lambda^{2}-1)\,
	\delta^{[a}_{[b}\,\delta^{c]}_{d]}\,,
\ee
where $L^{a}{}_{b}$ are the generators of $\mathfrak{sl}_{4}$
with $a,b=1,\ldots,4$ and $L^{a}{}_{a}=0$\,.
In fact, these elements
generate an ideal in the UEA of $\mathfrak{sl}_{N}$\ for any $N$\,.
In particular for $\mathfrak{sl}_2$\,, the $3D$ HS algebra $hs[\lambda]$
can be obtained in this way.
We will come back to this point later.

\subsection{Minimal representations}
\label{sec: min}

In order to obtain HS algebra from the UEA,
we quotient the UEA with an ideal corresponding
to the tensors which are not of the Killing-tensor type \eqref{M yd}.
This quotienting procedure fixes all the Casimir operators
as well as the underlying representation:
the ideal coincides with the kernel of such representations.
Indeed, this reduction of generators can be carried out by simply choosing the proper representation of the isometry algebra,
which is small enough to project all the
generators except for Killing tensors.
It turns out that these representations are in fact the \emph{smallest} ones,
namely the minimal representations \cite{Joseph:1974}.
Here, for the self-completeness, we provide a brief introduction
to the minimal representations
by mainly focusing on the case of the classical Lie algebras over $\mathbb C$\,.

Let us first introduce the convention which we shall adopt in the following discussion:
\begin{itemize}
\item
Let us begin with $\mathfrak{sp}_{2N}$ which is generated by elements $N_{AB}$\,:
\be
	N_{[AB]}=0\,,\qquad A,B=1,2,\ldots,2N\,,
\ee
with the commutation relation,
\be
[\![\,N_{AB}\,,\ N_{CD}\,]\!]=
	\Omega_{A(C}\,N_{D)B}+\Omega_{B(C}\,N_{D)A}\,.
\ee
Here, $\Omega_{AB}=-\Omega_{BA}$ is the symplectic matrix,
with  the inverse $\Omega^{AB}$\,:
\be
	\Omega^{AB}\,\O_{BC}
	=\O_{CB}\,\O^{BA}=\delta^{A}_{C}\,,
\ee
which is used to lower the indices as $V_{A}=\Omega_{AB}\,V^{B}$\,.
\end{itemize}
Now, we move to $\mathfrak{sl}_{N}$ and $\mathfrak{so}_{N}$ which we shall
describe as subalgebras of $\mathfrak{sp}_{2N}$\,.
For that, it is convenient to organize the $\mathfrak{sp}_{2N}$ indices $A$ as
\be
	A=\a\,a\,,\qquad \a=\pm\,,\quad a=1,2,\ldots, N\,,
\ee
with which the symplectic matrix becomes
\be
	\Omega_{\a a\:\b b}=\epsilon_{\a\b}\,\eta_{ab}\,,
	\qquad
	\epsilon_{\pm\mp}=\pm1\,.\label{omega}
\ee
\begin{itemize}
\item
The $\mathfrak{sl}_{N}$ is generated by
the traceless elements \mt{L^{a}{}_{b}:=N_{-}{}^{a}{}_{\:+b}-\frac1N\,
\delta^{a}_{b}\,N_{-}{}^{c}{}_{\:+c}}
with the commutation relation,
\be
	[\![\,L^{a}{}_{b}\,,\ L^{c}{}_{d}\,]\!]=
	\d^{c}_{b}\,L^{a}{}_{d}-\d^{a}_{d}\,L^{c}{}_{b}\,.
\ee
\item
The $\mathfrak{so}_{N}$ is generated by
the antisymmetric elements \mt{M_{ab}:=N_{-a\:+b}-N_{-b\:+a}}
with the commutation relation \eqref{so CR}\,.
\end{itemize}
The dual vector space of these Lie algebras
are the spaces of matrices $U^{AB}$, $V_{a}{}^{b}$ and $W^{ab}$
with $U^{[AB]}=0$\,, $V_{a}{}^{a}=0$ and $W^{(ab)}=0$\,.
It is convenient to introduce
\be
	N(U)=\frac12\,N_{AB}\,U^{AB}\,,\qquad
	L(V)=L^{a}{}_{b}\,V_{a}{}^{b}\,,\qquad
	M(W)=\frac12\,M_{ab}\,W^{ab}\,,
\ee
in terms of which the commutation relations of the algebras can be also given by
\be
	[\![\,T(A_{1})\,,\,T(A_{2})\,]\!]
	=T(A_{1}\,A_{2}-A_{2}\,A_{1})\,.
\ee
Here, $(T, A)$ are $(N,U), (L,V)$ or $(M,W)$\,,
while the products of the dual matrices are given by
$(U_{1}\,U_{2})^{AB}={U_{1}}^{AC}\,\Omega_{CD}\,{U_{2}}^{DB}$\,,
$(V_1\,V_2)_a{}^{b}={V_{1}}_{a}{}^{c}\,{V_{2}}_{c}{}^{b}$ and
$(W_1\,W_2)^{ab}={W_1}^{ac}\,\eta_{cd}\,{W_2}^{db}$\,.

\medskip

Now let us come back to the introduction to minimal representations
for classical Lie algebras.
There are several different approaches to minimal representations.\footnote{See e.g.
 \cite{Joseph:1976,Garfinkle:1982,Levasseur:1988,Binegar:1991,Braverman:1998,Li:2000,Gan:2004,Eastwood:2007,Fronsdal:2009,Todorov:2010md,Kobayashi:2013ao} and references therein for general introduction to minimal representation.}
 Here, we follow the coadjoint orbit method
 where minimal representation is given as
 the quantization of  the minimal nilpotent orbit.
The coadjoint action of a Lie group $\mathcal G$ on the dual space $\mathfrak g^{*}$ of
its Lie algebra $\mathfrak g$ is defined by
\be
	({\rm Coad}_{g}\,A)(T) = A(g^{-1}\,T\,g)\,,
\ee
where $g, T$ and $A$ are respectively any elements of
$\mathcal G, \mathfrak{g}$ and $\mathfrak{g}^{*}$\,.
Each orbit under such actions
--- coadjoint orbit --- is an even dimensional subspace of $\mathfrak g^{*}$
with $\mathcal G$-invariant symplectic form.
While there exists a continuum of semi-simple orbits,
the number of nilpotent orbits is finite.
The semi-simple orbits and the principal nilpotent orbit
--- the unique dense orbit of the nilpotent orbits --- are
given by a set of equations involving the dual of Casimir operators.
Hence, their dimension is
\mt{{\rm dim}\,\mathfrak  g -{\rm rank}\,\mathfrak  g}\,.
The other nilpotent orbits have smaller dimensions
as they are defined by a larger number of polynomial equations.
The nilpotent orbit with minimum dimension --- apart from
the trivial orbit $\{0\}$ --- is also unique and called the minimal orbit
$\mathcal O_{\rm min}(\frak g)$\,, which is what we are interested in.
The minimal orbits of classical Lie algebras are
determined by the following quadratic equations \cite{Fronsdal:2009}\,:
\ba
	&\mathcal O_{\rm min}(\frak{sp}_{2N})\ &:\quad
	U^{A[B}\,U^{D]C}=0\,,\nn
	&\mathcal O_{\rm min}(\frak{sl}_{N})\ &:\quad
	V_{[a}{}^{b}\,V_{c]}{}^{d}=0\,,
	\label{min orbit} \\
	&\mathcal O_{\rm min}(\frak{so}_{N})\ &:\quad
	W^{a[b}\,W^{cd]}=0=W^{ab}\,W_{b}{}^{c}\,,\nonumber
\ea
and can be parameterized by
\ba
	U^{AB} \eq u^{A}\,u^{B}\,,\nn
	V_{a}{}^{b} \eq v_{+a}\,v_{-}^{b} \qquad [v_{+}\!\cdot v_{-}=0]\,,
	\label{min para}\\
	W^{ab} \eq w_{+}^{[a}\,w_{-}^{b]}\qquad [w_{\a}\!\cdot w_{\b}=0]\,.
	\nonumber
\ea
From the above, one can deduce the dimensions of these minimal orbits as
\begin{center}
    \begin{tabular}{ | c | c | c | c | c |}
    \hline
  	$\mathfrak{g}$ & ${\rm dim}\,\mathfrak g$ &
	${\rm rank}\,\mathfrak g$ &
	${\rm dim}\,\mathcal O_{\rm prin}(\frak g)$ & $
	{\rm dim}\,\mathcal O_{\rm min}(\frak g)$
	\\ \hline\hline
	$\mathfrak{sp}_{2N}$ & $N(2N+1)$ & $N$ & $2\,N^{2}$ & $2N$
	\\ \hline
	$\mathfrak{sl}_{N}$ & $N^{2}-1$ & $N-1$ & $N(N-1)$ & $2(N-1)$
	\\ \hline
	$\mathfrak{so}_{N}$ & $\frac{N(N-1)}2$ & $\big\lfloor\frac N2\big\rfloor$ & $\big\lceil\frac {N(N-2)}2\big\rceil$ & $2(N-3)$
	\\ \hline
    \end{tabular}
\end{center}
where
$\lfloor x\rfloor={\rm max}\{m\in \mathbb Z\,|\,m\le x\}$ and
$\lceil x\rceil={\rm min}\{n\in \mathbb Z\,|\,n\ge x\}$\,.
The kernel of the minimal representation
in the UEA  is the Joseph ideal $\cal J(\frak g)$
(the characteristic variety of the Joseph ideal
is the closure of the minimal orbit $\cal O_{\rm min}(\frak g)$).
Joseph ideal is the ideal of UEA
generated by certain elements in \mt{\mathfrak{g}\odot \mathfrak{g}}\,.
In the case of classical Lie algebras,
one can equivalently consider the ideals of the tensor algebra
generated by the relations \cite{Fronsdal:2009}\,:
\ba
	&\mathcal J(\frak{sp}_{2N})\ &:\quad
	N_{A[B} \otimes N_{C]D}+\frac{\hbar}{2}
	\left(\Omega_{A[B}\,N_{C]D}+\Omega_{D[B}\,N_{C]A}
                         -\Omega_{BC}\,N_{AD}\right)\nn
	&&\hspace{84pt}+\,\frac{\hbar^2}{2}
	\left(\Omega_{A[B}\,\Omega_{C]D}
                         -\Omega_{BC}\,\Omega_{AD}\right)\sim 0\,,\nn
	&\mathcal J(\frak{sl}_{N})\ &:\quad
	L^{[a}{}_{b} \otimes L^{c]}{}_{d}
	+\hbar\left(\delta^{[a}_{(\beta}\,L^{c]}{}_{d)} +
	\lambda\,\delta^{[a}_{[b}\, L^{c]}{}_{d]}\right)+
	\hbar^{2}\,\frac{\lambda^{2}-1}4\,
	\delta^{[a}_{[b}\,\delta^{c]}_{d]}\sim 0\,,
	\label{Joseph}\\
	&\mathcal J(\frak{so}_{N})\ &:\quad
	M_{a[b} \otimes M_{cd]}-\hbar\,\eta_{a[b}\,M_{cd]}
	\sim 0\sim
	M_{c(a} \otimes M_{b)}{}^{c}-\hbar^{2}\,\frac{N-4}2\,\eta_{ab}
	\nonumber\,,
\ea
which are dual analog of the ones \eqref{min orbit}
on $\frak{g}^{*}$\,.
Here, we have introduced the deformation parameter $\hbar$
--- which are taken as \mt{\hbar=1} in the rest of the paper ---
in order to make manifest that the above relations provide quantizations of the orbits
given in \eqref{min orbit}.
Inversely, an irreducible representation of $\frak g$
associated with a coadjoint orbit can be obtained by quantizing
the orbit.
Notice also that for the $\frak{sl}_{N}$ series,
the quantization or equivalently the minimal representation is not unique
but is of one-parameter family labeled by $\lambda$\,.
For the $\frak{sl}_{4}$ case, the minimal representation
has been studied also in \cite{Fernando:2009fq}.

Let us note that
the minimal representations are the
irreducible representations with the minimum non-zero
Gelfand-Kirillov (GK) dimension \cite{GK}.
The GK dimension of a vector space can be understood roughly as
the number of continuous variables necessary
to describe the vector space.
In case of the minimal representation,
it is half of the dimension of minimal orbit.
For one-particle Hilbert space, it corresponds to the dimension
of the space --- not the spacetime --- where the wave function lives.
Several conclusions can be drawn from this perspective:
\begin{itemize}
\item
A particle in $D\ge4$ dimensions has GK dimension $D-1$\,.
Collecting finitely many particles
cannot increase the GK dimension,
since it amounts to introducing some finite-range discrete variables which
label the particles.
On the contrary, an infinite collection of particles may have a bigger GK dimension: for example, Kaluza-Klein compactification decomposes
a particle in higher dimensions --- therefore, of higher GK dimension ---
into an infinite set of particles in lower dimensions.

\item
$(A)dS_{D}$ has isometry (a real form of) $\mathfrak{so}_{D+1}$\,,
whose minimal representation has GK dimension $D-2$\,. Hence,
a particle in $(A)dS_{D}$ cannot be minimal but
a representation associated with the next-to-minimal orbit
of $\mathfrak{so}_{D+1}$\,.

\item
The particles corresponding to conformal fields, namely singletons, in $d$ dimensions have  GK dimension $d-1$\,,
the same as the minimal representation of the
conformal algebra $\mathfrak{so}_{d+2}$\,.
However, apart from the scalar, the other conformal-field representations
require additional helicity labels.
The spinor in \mt{d=3} and helicity-$h$ representations in \mt{d=4}
are exceptions as they have only one helicity component.
Actually, such representations underlie
the $D=4$ and $D=5$ HS algebras, respectively:
in the former case the scalar and spinor singletons, namely Rac and Di,
give the same HS algebra, while
in the latter case the helicity $h$ is related to the $\lambda$ deformation \eqref{lambda J} by $h=\lambda$\,.

\item
The Flato-Fronsdal theorem \cite{Flato:1978qz} --- as well as its generalization \cite{Vasiliev:2004cm} ---
corresponding to the HS $AdS_{d+1}/CFT_{d}$ duality  can be also
understood in this way. If the conformal field theory on the boundary has
a finite content of fields, then its GK dimension
is $d-1$\,. The space of bilinear operators,
which corresponds to the tensor product of two singletons,
has doubled GK dimension
which can be viewed as
\be
	2(d-1)=d+(d-2)\,.
\ee
Here, $d$ is the number of space variables of $(A)dS_{d+1}$\,,
and \mt{d-2} is
the GK dimension corresponding to the
helicity variables.
For instance, in the scalar singleton case, the corresponding dual
$(A)dS_{d+1}$ fields are massless symmetric fields of all integer spins.
The number of their helicity states up to spin $s$
is $\sim s^{d-2}$\,, from which we can deduce the
corresponding GK dimension.

\item
Finally, suppose we consider a $D$-dimensional field theory with a finite
field content which carries a representation of a global-symmetry
Lie algebra $\mathfrak{g}$\,.
Then, the GK dimension of the Hilbert space cannot be smaller than
that of the minimal representation of $\mathfrak{g}$\,:
\be
	D-1\ge\frac12\,{\rm dim}(\mathcal{O}_{\rm min}(\mathfrak{g}))\,.
	\label{finite admc}
\ee
Among classical Lie algebras, $\mathfrak{sl}_{N}$ with
\mt{N\ge D+1},
$\mathfrak{so}_{N}$ with \mt{N\ge D+3}
and $\mathfrak{sp}_{2N}$ with \mt{N\ge D}
are already excluded with this condition.
Together with the requirement that $\mathfrak{g}$ contains
the $D$-dimensional isometry algebra $\mathfrak{so}_{D+1}$,
one gets much stronger restrictions: for instance, the only possible
orthogonal algebras $\mathfrak g$ are
$\mathfrak{so}_{D+1}$ itself and $\mathfrak{so}_{D+2}$\,.\footnote{In
$3D$, massless HS particles have GK dimension 1 corresponding
to the would-be gauge mode on the asymptotic boundary,
and the global symmetry is rather
the asymptotic symmetry than the bulk isometry one.
In case of $\mathfrak{sl}_{N}\oplus\mathfrak{sl}_{N}$
Chern-Simons theory,
the asymptotic symmetry is given by
$\mathcal W_{N}\oplus \mathcal W_{N}$
\cite{Campoleoni:2010zq}.
Interestingly, $\mathcal W_{N}$ does not contain $\mathfrak{sl}_{N}$
as subalgebra --- at least, not manifestly. If it did, the
GK dimension of the Hilbert space would be bigger or equal to
 $N-1$\,, which is not the case for $N\ge3$\,.

Let us note that in order to derive \eqref{finite admc},
we have required  the Hilbert space to carry a representation of
the global symmetry.
As we have seen in the $3D$ case, the presence of the asymptotic boundary may provoke a deformation of global symmetry
invalidating this condition.
This phenomenon is in principle possible in any odd $D$ dimensions,
so may provide a chance
for a consistent HS theory with a finite field content ---
see \cite{Manvelyan:2013oua} for a recent attempt.}

\end{itemize}

\subsection{HS algebras and Reductive dual pairs}
\label{sec: def}

The HS algebra defined in Section \ref{sec: UEA} is the symmetry (algebra) of the minimal representation of $\mathfrak{so}_{N}$\,.
Abusing this terminology to the other Lie algebras,
let us consider \emph{HS algebra} associated with a Lie algebra $\mathfrak g$\,:
\be
	hs(\mathfrak{g})=\cal U(\mathfrak{g})/ \cal J(\mathfrak{g})\,,
\ee
where $\cal U(\mathfrak{g})$
and $\cal J(\mathfrak{g})$
are respectively the UEA
and Joseph ideal of $\mathfrak g$\,.
As a vector space, $hs(\mathfrak{g})$ corresponds to  the space of polynomials in
$\cal O_{\rm min}(\frak g)$\,:
\be
	N(U)=\sum_{n=0}^{\infty} N^{\sst (n)}(U)\,,
	\qquad
	L(V)=\sum_{n=0}^{\infty} L^{\sst (n)}(V)\,,
	\qquad
	M(W)=\sum_{n=0}^{\infty} M^{\sst(n)}(W)\,,
	\label{gen fn}
\ee
with
\ba
	N^{\sst (n)}(U) \eq
	\frac1{2^{n}\,n!}\,N_{A_{1}B_{1},\ldots, A_{n} B_{n}}\,
	U^{A_{1}B_{1}}\cdots
	U^{A_{n}B_{n}}\,,
	\label{N n}\\
	L^{\sst (n)}(V) \eq
	\frac1{n!}\,L^{a_{1}\,\cdots\,a_{n}}_{b_{1}\,\cdots\, b_{n}}\,
	V_{a_{1}}{}^{b_{1}}\cdots
	V_{a_{n}}{}^{b_{n}}\,,
	\label{L n}\\
	M^{\sst(n)}(W) \eq
	\frac1{2^{n}\,n!}\,M_{a_{1}b_{1},\ldots, a_{n} b_{n}}\,
	W^{a_{1}b_{1}}\cdots
	W^{a_{n}b_{n}}\,.
	\label{M n}
\ea
Therefore, the expansion coefficients,
\be
N_{A_{1}B_{1},\ldots, A_{n} B_{n}}\,,
\qquad L^{a_{1}\,\cdots\,a_{n}}_{b_{1}\,\cdots\, b_{n}}\,,
\qquad M_{a_{1}b_{1},\ldots, a_{n} b_{n}}\,,
\label{gen HS}
\ee
are the generators of
$hs(\frak{sp}_{2N})$\,,
$hs_{\lambda}(\frak{sl}_{N})$
and $hs(\frak{so}_{N})$, respectively.
Due to the properties of minimal orbits \eqref{min orbit},
these generators can be chosen to be traceless:
\be
	\Omega^{A_{1}A_{2}}\,N_{A_{1}B_{1},\ldots, A_{n} B_{n}}
	=0\,,\quad\ \
	\delta^{b_{1}}_{a_{1}}\,
	L^{a_{1}\,\cdots\,a_{n}}_{b_{1}\,\cdots\, b_{n}}=0\,,
	\quad\ \
	\eta^{a_{1}a_{2}}\,M_{a_{1}b_{1},\ldots, a_{n} b_{n}}=0\,.
\ee
We will use the symbol $\lstar$ for the product of $hs(\mathfrak{g})$\,,
which is defined by
 \mt{\lstar\!:=\otimes/\!\sim}\,.
Here, $\sim$ is the equivalence relation \eqref{Joseph}.

For a classical Lie algebra $\frak g$\,,
instead of using the explicit form of Joseph ideals,
one can rely on the reductive dual pairs  to handle the algebraic structure of
$hs(\mathfrak{g})$\,:
see \cite{Sezgin:2001zs,Vasiliev:2001wa} and \cite{Vasiliev:2003ev}
for the $\mathfrak{sl}_{4}$ and $\mathfrak{so}_{D+1}$ cases, respectively,
and  for more generalities  see e.g. \cite{Li:2000,Todorov:2010md} and references therein.
A reductive dual pair in the symplectic group $Sp_{2N}$
 is a pair of subgroups,
 \be
 	(\,G_{1}\,,\,G_{2}\,)\subset Sp_{2N}\,,
\ee
which are centralizers of each other.
Then, there is a bijection between two irreducible representations
$\pi_{1}$ and $\pi_{2}$ of $G_{1}$ and $G_{2}$
so that
for any $\pi_{1}$ (or $\pi_{2}$) there exists
at most one $\pi_{2}$ (or $\pi_{1}$).
The minimal representations
of $\mathfrak{sl}_{N}$ and $\mathfrak{so}_{N}$ can be obtained
from that of $\mathfrak{sp}_{2N}$ by
considering the dual pairs,
\be
	(\,G_{1}\,,\,G_{2}\,)=(\,GL_{1}\,,\,GL_N\,)\quad
	{\rm and}
	\quad
	(\,Sp_{2}\,,\,O_{N}\,)\,,
\ee
respectively.
For the former case,
we take the  representation
of $GL_{1}$  labeled by $\lambda$
--- so we can see again that the minimal representation of $\mathfrak{sl}_{N}$
has one-parameter family.
For the latter case,
we take the trivial representation of $Sp_{2}$\,.
In the following,
we review how one can deal
with the explicit structures of HS algebras
using such dual pair correspondences.

\paragraph{$\bm{hs(\frak{sp}_{2N})}$}
Notice first that the minimal representation of $\frak{sp}_{2N}$
is the metaplectic representation
described by oscillators $y_{A}$\,:
\be
	N_{AB}=y_{A}\,y_{B}\,,
\ee
endowed with
the Moyal $\star$ product,
\be\label{star sp}
	(f\star g)(y)=
	\exp\left(\,\frac12\,\Omega_{AB}\,\partial_{y_{A}}\,\partial_{z_{B}}
	\right)
	f(y)\ g(z)\,\Big|_{z=y}\,.
\ee
Hence, $hs(\frak{sp}_{2N})$ is generated by polynomials
of $y_{A}\,y_{B}$,
that is,
the space of all even-order polynomials in $y_{A}$:
\be
	N_{A_{1}B_{1},\ldots, A_{n} B_{n}}
	=y_{A_{1}}\,y_{B_{1}}\,\cdots\,y_{A_{n}}\,y_{B_{n}}\,,
\ee
and the generating function $N(U)$ \eqref{gen fn} becomes a Gaussian,
\be\label{gen fn N}
	N(U)=\exp\left(\frac12\ y_{A}\,U^{AB}\, y_{B}\right).
\ee
In this case, the product of  $hs(\frak{sp}_{2N})$ coincides with the Moyal product:
$\lstar\!=\star$\,.

\paragraph{$\bm{hs_{\lambda}(\frak{sl}_{N})}$}

We consider the $\mathfrak{gl}_{1}$-center of
$hs(\frak{sp}_{2N})$\,, that is, the set of elements satisfying
\be
	\big[\,y_{+}\!\cdot y_{-}\,,\ f(y)\,\big]_{\star}
	=\left(y_{-}\cdot\partial_{y_{-}}-y_{+}\cdot\partial_{y_{+}}\right)f(y)=0\,.
\ee
The solution space is generated by
\be
	\tilde L^{a_{1}\,\cdots\,a_{n}}_{b_{1}\,\cdots\,b_{n}}
	=y_{-}^{\ a_{1}}\,y_{+b_{1}}\cdots\,y_{-}^{\ a_{n}}\,y_{+b_{n}}\,,
	\label{tilde L}
\ee
whose traceless part can be identified with the generator $L^{a_{1}\,\cdots\,a_{n}}_{b_{1}\,\cdots\,b_{n}}$ of $hs_{\lambda}(\frak{sl}_N)$\,.
It is convenient for later use to generalize the definitions \eqref{gen fn} and \eqref{L n} to $\tilde L^{a_{1}\,\cdots\,a_{n}}_{b_{1}\,\cdots\,b_{n}}
$ getting
\be
	\tilde L(\tilde V)
	=\exp\left(y_{-}\!\cdot\tilde V\cdot y_{+}\right),
	\label{gen L tilde}
\ee
where the matrix-variable $\tilde V^{b}_{a}$ satisfies	
\be
	\tilde V_{[a}{}^{b}\,\tilde V_{c]}{}^{d}=0
	\quad \Leftrightarrow \quad
	\tilde V_{a}{}^{b}=\tilde v_{+a}\,\tilde v_{-}^{b}\,.
	\label{tilde V}
\ee
This space is also endowed with the $\star$ product,
and we will refer to this algebra as $hs(\mathfrak{gl}_{N})$\,.
In order to get the HS algebra of $\mathfrak{sl}_{N}$\,,
we take an irreducible representation of $\mathfrak{gl}_{1}$\,,
and this amounts to quotienting $hs(\mathfrak{gl}_{N})$ by the relation,
\be
	K_{\lambda}:=y_{+}\!\cdot y_{-}-\frac{N}{2}\,\lambda\sim 0\,.
	\label{K rel}
\ee
Let us make a brief remark here: consider,
before taking the quotient by $K_{\lambda}$\,,
the following isomorphism:
\ba
	\rho_{\l}\ :\ hs(\mathfrak{gl}_{N})\quad &\to& \quad hs_{\l}(\mathfrak{gl}_{N})\,,\nn
	f(y) \quad \quad &\mapsto & \quad \rho_{\l}(f)(y)=e^{\tfrac{\l}{2}\,\partial_{y_{+}}\!\cdot\,\partial_{y_{-}}}\,
	f(y)\,.
	\label{alg mor}
\ea
Then, the image $hs_{\l}(\mathfrak{gl}_{N})$ admits a deformed $\star$ product,
\ba
	&& (f\star_{\lambda} g)(y) = \rho_{\l}\big(\,\rho^{-1}_{\l}(f)
	\star \rho^{-1}_{\l} (g)\,\big)(y)\nn
	&&=
	 \exp\left[\,\frac12\left(\partial_{y_{+}}\!\cdot\partial_{z_{-}}-\partial_{z_{+}}\!\cdot\partial_{y_{-}}\right)
            +\frac{\l}{2}\left(\partial_{y_{+}}\!\cdot\partial_{z_{-}}+\partial_{z_{+}}\!\cdot\partial_{y_{-}}\right)
	\right]
	f(y)\ g(z)\,\bigg|_{z=y}\,.
	\label{starl}
\ea
The algebra $hs_{\lambda}(\frak{sl}_{N})$ can be
equivalently obtained by quotienting $hs_{\lambda}(\frak{gl}_{N})$
by the relation,
\be
	\rho_{\l}(K_{\lambda})
	=K_{0}\sim  0\,.
\ee
Hence, although $hs_{\lambda}(\frak{gl}_{N})$ are all equivalent for different $\l$,
after quotienting, they become distinct algebras  $hs_{\lambda}(\frak{sl}_{N})$\,.
In the following, we shall use the description
$hs_{\lambda}(\frak{sl}_{N})=hs(\frak{gl}_{N})/\la K_{\lambda}\sim 0 \ra$
for explicit computations.

An element of $hs_{\lambda}(\frak{sl}_{N})$ is a class representative
 $[\![\,a\, ]\!]$
 for elements $a \in hs(\frak{gl}_{N})$,
and the product of $hs_{\lambda}(\frak{sl}_{N})$ is defined by
\be
	[\![ \,a\, ]\!] \lstar [\![ \,b\, ]\!]
	:= [\![ \,a\star b\, ]\!]\,.
\ee
In order to get an explicit expression for the product $\!\lstar\!$\,,
we need to  choose a class representative
for a generic element of $hs(\frak{gl}_{N})$\,.
We can do this for $\tilde L^{\sst (n)}$,
and for concreteness let us consider $\tilde L^{\sst (2)}$\,,
which can be decomposed into  $L^{\sst (n)}$ with $n=0,1,2$ as
\be
	\tilde L^{\sst (2)}(\tilde V)
	=L^{\sst (2)}(\tilde V)
	+\frac{4}{N+2}\ y_{+}\!\cdot y_{-}\ \tilde v_{+}\!\cdot \tilde v_{-}\,L^{\sst (1)}(\tilde V)
	+\frac{2}{N(N+1)}(\,y_{+}\!\cdot y_{-})^{2}\,(\,\tilde v_{+}\!\cdot \tilde v_{-})^{2}\,.
	\label{trace decomp}
\ee
From this example, one can notice that $\tilde L^{\sst (n)}$
and $L^{\sst (n)}$ do not belong to the same equivalence class
since the trace part of $\tilde L^{\sst (n\ge2)}$ cannot be
written as \mt{K_{\lambda}\star a}
 for an element $a\in hs(\frak{gl}_{N})$\,.
We can remove all the \mt{y_{+}\!\cdot y_{-}} terms in
the traceless decomposition of $\tilde L^{\sst (n)}$
using $\star$ product and the relation \eqref{K rel}.
Since this procedure is unambiguous, it can be served to choose
a class representative.
For $n=2$ case, we get
\be
	\tilde L^{\sst (2)}(\tilde V)
	\sim L^{\sst (2)}(\tilde V)
	+\frac{2N\l}{N+2}\,\tilde v_{+}\!\cdot \tilde v_{-}\ L^{\sst (1)}(\tilde V)
	+\frac{N\l^2+1}{2(N+1)}(\,\tilde v_{+}\!\cdot\tilde v_{-})^{2}=\big[\hspace{-3pt}\big[\,
	\tilde L^{\sst (2)}(\tilde V)
	\,\big]\hspace{-3pt}\big]\,.
\ee
In general, the class representative of $\tilde L^{\sst (n)}$
has the following form of series:
\be
	\big[\hspace{-3pt}\big[\,
	\tilde L^{\sst (n)}(\tilde V)
	\,\big]\hspace{-3pt}\big]
	=\sum_{m=0}^{n}\,
	s^{\sst(n)}_{m}\,\frac{\la\tilde V\ra^{m}}{m!}\,L^{\sst (n-m)}(\tilde V)\,,
	\label{rep L}
\ee
where coefficients $s^{\sst(n)}_{m}$ are fixed ones, in principle calculable, but their explicit expressions are not necessary for our purpose.
Let us comment that in the above series
only the structure $\la \tilde V\ra^{m}$ can appear as the coefficient of
$L^{\sst (n-m)}$ since it is the unique $m$-th order scalar in $\tilde V$
due to the property \eqref{tilde V}.

\paragraph{$\bm{hs(\frak{so}_{N})}$}

Similarly to the $\frak{sl}_{N}$ case, we first consider
the $\frak{sp}_{2}$ center of $hs(\frak{sp}_{2N})$\,, that is, the set of elements satisfying
\be
	\big[\, y_{\a}\!\cdot y_{\b}\,,f(y)\,\big]_{\star}
	=\left(y_{\a}\cdot\partial_{y^{\b}}+y_{\b}\cdot\partial_{y^{\a}}\right)f(y)=0\,.
\ee
The solution space is again endowed with the $\star$ product, and
we refer to this algebra as $\widetilde{hs}(\frak{so}_{N})$\,.
It is generated by
\be
	\tilde M_{a_{1}b_{1}\,\cdots\,a_{n}b_{n}}=
	2^n\,y_{[-a_{1}}\,y_{+]b_{1}}\,\cdots\,y_{[-a_{n}}\,y_{+]b_{n}}\,,
\ee
and we identify its traceless part with the generators $M_{a_{1}b_{1}\,\cdots\,a_{n}b_{n}}$ of $hs(\frak{so}_N)$.
Again generalizing the definitions \eqref{gen fn} and \eqref{M n}
to $\tilde M_{a_{1}b_{1}\,\cdots\,a_{n}b_{n}}$\,, we get
\be
	\tilde M(\tilde W)
	=\exp\left(\,y_{-a}\,\tilde W^{ab}\,y_{+b}\right),
	\label{gen M tilde}
\ee
with the matrix-variable $\tilde W$ satisfying
\be
	\tilde W^{a[b}\,\tilde W^{cd]}=0\quad \Leftrightarrow
	\quad
	\tilde W^{ab}=\tilde w_{+}^{[a}\,\tilde w_{-}^{b]}\,.
	\label{tilde W}
\ee
The HS algebra, $hs(\frak{so}_{N})$\,,
is the quotient of $\widetilde{hs}(\frak{so}_{N})$ by the relation,
\be
	K_{\a\b}:=y_{\a}\cdot y_{\b}\sim 0\,,
	\label{so K}
\ee
which corresponds to taking the trivial representation of $\mathfrak{sp}_{2}$\,.
The class representative is given, analogously to
$hs_{\l}(\mathfrak{sl}_{N})$\,, by a series,
\be
	\big[\hspace{-3pt}\big[\,
	\tilde M^{\sst (n)}(\tilde W)
	\,\big]\hspace{-3pt}\big]
	=\sum_{m=0}^{[n/2]}\,t^{\sst(n)}_{m}
	\frac{\la\tilde W^{2}\ra^{m}}{m!}
	\,M^{\sst (n-2m)}(\tilde W)\,.
	\label{rep M}
\ee
Remark that the structure $\la\tilde W^{2}\ra^{m}$
in front of $M^{\sst (n-2m)}$
is again the unique possibility due to the property \eqref{tilde W}.

\section{Trace and Structure constants of HS algebras}
\label{sec: bt}

In this section, we shall derive explicit form of the structure constants
of the previously defined HS algebras associated with classical Lie algebras.
Let us begin with recalling that the structure constant $C_{\bm{ab}}{}^{\bm c}$
of HS algebra $hs(\mathfrak g)$ is defined by
\be
	T_{\bm a}
	\lstar T_{\bm b}
	= C_{\bm{ab}}{}^{\bm c}\;T_{\bm c}\,,
	\label{str cnst}
\ee
where $T_{\bm a}$
is one of the generators \eqref{gen HS} of $hs(\mathfrak g)$\,,
and $\bm a, \bm b, \bm c$ are the collective indices.
A convenient way to address structure constant is
by making use of \emph{trace} of HS algebra,
defined
as the identity piece of given element --- see e.g. \cite{Vasiliev:2004cm} for more details:
\be
	\tr\left[\,c_{0}+c^{\bm a}\,T_{\bm a}\right]=c_{0}\,.
\ee
From the existence of the antiautomorphism
\mt{T^{\sst (n)} \mapsto (-1)^{n}\,T^{\sst (n)}},
one can show that the bilinear form,
\be
	B_{\bm {ab}}
	=\tr\left[\,T_{\bm a} \lstar T_{\bm b}\,\right],\label{bilinear form}
\ee
is symmetric and invariant.
The trilinear form is simply related to the structure constant and the bilinear form as
\be
	C_{\bm {abc}}
	=\tr\left[\,T_{\bm a}\lstar T_{\bm b} \lstar  T_{\bm c}\,\right]
	=C_{\bm{ab}}{}^{\bm d}\,\,
	B_{\bm{dc}}\,.
\ee
In the notation introduced in the previous section,
the trace is given simply by
\be
	\tr\left[\,T(A)\,\right]=T(0)\,,
	\label{trace}
\ee
while the bilinear and trilinear forms read
\ba
	\mathcal B(A_{1},A_{2})\eq \tr\left[\,T(A_{1})\lstar T(A_{2})\,\right],\nn
	 \mathcal C(A_{1},A_{2},A_{3}) \eq
	 \tr\left[\,T(A_{1})\lstar T(A_{2})\lstar T(A_{3})\,\right].
	 \label{bi tri}
\ea
In the following, for each of $hs(\mathfrak{sp}_{2N})$\,,
$hs_{\lambda}(\mathfrak{sl}_{N})$ and $hs(\mathfrak{so}_{N})$\,,
we shall work out the trace
and the bi-/trilinear forms.
For $hs(\mathfrak{so}_{N})$ and $hs_{\lambda}(\mathfrak{sl}_{2})$\,
the bilinear forms have been obtained respectively in \cite{Vasilev:2011xf} and \cite{Vasiliev:1989re}.
Let us remark as well that
the structure constants of $hs(\mathfrak{so}_{5})$  and $hs_{\lambda}(\mathfrak{sl}_{2})$
have been proposed respectively in \cite{Fradkin:1986ka} and \cite{Pope:1989sr,Pope:1990kc,Fradkin:1990ir}.

Actually, in the case of $hs(\mathfrak{sp}_{2N})$\,, the algebraic structure
is already explicit at the level of $\star$ product since there is no
quotienting process to perform.
However, we decide to treat the algebras $hs(\mathfrak{sp}_{2N})$\,, $hs_{\l}(\mathfrak{sl}_{N})$ and  $hs(\mathfrak{so}_{N})$ in the equal footing,
for the sake of remarking the similar algebraic properties they possess and making manifest the relations between them.

\subsection{$hs(\mathfrak{sp}_{2N})$}
\label{sec: sp}

Let us begin with $hs(\mathfrak{sp}_{2N})$\,,
whose essential ingredients can be found e.g. in \cite{Didenko:2003aa}.

\paragraph{Trace}

In this case, the trace defined as \eqref{trace} is equivalent simply to
\be
	\tr\left[f(y)\right]=f(0)\,,
	\label{sp trace}
\ee
for an element $f(y)$ of the algebra $hs(\mathfrak{sp}_{2N})$\,.

\paragraph{Structure constant}

Let us consider the bilinear and trilinear forms \eqref{bi tri}.
For that, we need to first evaluate the product of generating functions \eqref{gen fn N}
$N(U_{1})\lstar N(U_{2})$ and $N(U_{1})\lstar N(U_{2})\lstar N(U_{3})$\,.
Since $\lstar\!=\star$ for \mt{hs(\frak{sp}_{2N})}, we can rely on the composition property of the $\star$ product.
For the Gaussian functions of type,
\be
	\mathcal G(S)=
	\frac1{\sqrt{\det\left(\frac{1+S}2\right)}}\,\exp\left[y_{A}\left(\frac{S-1}{S+1}
	\right)^{\!\!AB} y_{B}\right],
	\label{Cayley def}
\ee
the $\star$ product admits a manifestly associative form:
\be
	\mathcal G(S_{1})\star \mathcal G(S_{2})=\mathcal G(S_{1}\,S_{2})\,.
    \label{Cayley comp}
\ee
The connection between $\mathcal G(S)$ and $N(U)$
 involves a Cayley transformation \cite{Didenko:2003aa}:
\be
	\mathscr C(U)=\frac{2+U}{2-U}\,,\qquad
	\mathscr C^{-1}(S)=2\,\frac{S-1}{S+1}\,,
	\label{Cayley transf}
\ee
and using the rule \eqref{Cayley comp}
and the trace formula \eqref{sp trace}, one gets
the $n$-linear forms as
\be
	\frac1{\sqrt{G^{\sst(n)}(U_{1},\ldots,U_{n})}}:=
	\tr\left[\,N(U_{1})\,\star\,\cdots\, \star\,N(U_{n})\,\right]\,,
\ee
where the function $G^{\sst (n)}$ is given by
\be
	G^{\sst(n)}(U)=
	\frac
	{\det_{2N}\left(\frac12\,
	\prod_{k=1}^{n}\frac{2+U_{k}}{2-U_{k}}+\frac12\right)}
	{\prod_{k=1}^{n}\det_{2N}\left(\frac12\,\frac{2+U_{k}}{2-U_{k}}+\frac12\right)}
	=
	\det{}_{\!2N}\bigg[\,\frac12\prod_{k=1}^{n}(1+U_{k})+\frac12\,\bigg]\,.
\ee
Here, for the second equality we have used the fact that $U_{k}^{2}=0$\,.
The $n=2$ case can be obtained immediately
using $U_{1}\,U_{2}\,U_{1}=\la U_{1}\,U_{2}\ra\,U_{1}$ as
\be
	G^{\sst (2)}(U_{1},U_{2})
	=1+\tfrac14\,\la\,U_{1}\,U_{2}\,\ra\,.
\ee
The $n=3$ case requires more calculations --- see Section \ref{sec: det} ---
and the result reads
\be
	G^{\sst (3)}(U)=
	1+\tfrac14\,\L(U)\,,
	\label{3 det sp}
\ee
where $\L(U)$ is defined by
\be
	\L(U):=\la\,U_{1}\,U_{2}\,\ra+\la\,U_{2}\,U_{3}\,\ra
	+\la\,U_{3}\,U_{1}\,\ra+
	\la\,U_{1}\,U_{2}\,U_{3}\,\ra\,.
\ee
Notice that, due to \mt{U_i^{AB}=U_i^{BA}} and \mt{\O_{AB}=-\O_{BA}}, the following identity is satisfied:
\be
	\la\,U_{1}\,U_{2}\,U_{3}\,\ra
	+\la\,U_{3}\,U_{2}\,U_{1}\,\ra=0\,.
\ee
Finally, we obtain the bilinear and trilinear forms as
\ba
	&&\mathcal B(U)=\tr[N(U_{1})\lstar N(U_{2})]=\frac1{\sqrt{1+\tfrac14\,\la\,U_{1}\,U_{2}\ra}}\,,\\
	&&\mathcal C(U)=\tr[N(U_{1})\lstar N(U_{2})\lstar N(U_{3})]=\frac1{\sqrt{1+\tfrac14\,\L(U)}}\,.
	\label{sp B C}
\ea
Notice that these bilinear and trilinear forms are given by the same function \mt{1/\sqrt{1+z}}\,, but with different arguments.

\subsection{$hs_{\lambda}(\mathfrak{sl}_{N})$}
\label{sec: sl}

Now we move to the $hs_{\lambda}(\mathfrak{sl}_{N})$ case,
where we need to handle the $\mathfrak{gl}_{1}$ coset.

\subsubsection*{Trace}

In order to conveniently deal with the coset structure,
we extend the definition of the trace \eqref{trace}
to $hs(\mathfrak{gl}_{N})$ supplementing it with the condition,
\be
	\tr\left(K_{\lambda}\star a\right)=0\,,\quad \forall a\in hs(\mathfrak{gl}_{N})\,,
\ee
so that we get
\be
	\tr(a_{1}\lstar\cdots \lstar a_{n})=\tr(a_{1}\star\cdots\star a_{n})\,,
	\qquad \forall a_{i}\in hs_{\l}(\frak{sl}_N)\,.
	\label{prod sl}
\ee
Since $a_{1}\star\cdots\star a_{n}$ belongs to
$hs(\mathfrak{gl}_{N})$\,,
we would like to have a trace formula for a generic element of $hs(\mathfrak{gl}_{N})$\,.
For that, we first consider the trace of the generating function $\tilde L(\tilde V)$ introduced in \eqref{gen L tilde}.
Using \eqref{rep L} and \eqref{trace}, we get
\be\label{trexp}
	\tr\big[\tilde L(\tilde V)\big]=\tr\big[\exp(y_{-}\!\cdot \tilde V\cdot y_{+})\big]
	=s(\la \tilde V\ra)\,,
	\qquad
	s(z)=\sum_{n=0}^{\infty}\,s^{\sst(n)}_{n}\,\frac{z^{n}}{n!}\,.
\ee
Hence, the trace of $\tilde L(\tilde V)$ is encoded in the function $s(z)$\,,
which requires the coefficients $s^{\sst(n)}_{n}$.
They can be obtained by taking the maximal trace of \eqref{rep L} as
\be\label{tn}
	s^{\sst (n)}_{n}=\frac{n!}{(N)_{n}}\,
	\big[\hspace{-3pt}\big[\,(y_{+}\!\cdot y_{-})^{n}
	\,\big]\hspace{-3pt}\big]\,,
\ee
where $(N)_n=N(N+1)\cdots (N+n-1)$ is the Pochhammer symbol.
The sequence
$\s_{n}:=[\![\,(y_{+}\!\cdot y_{-})^{n}]\!]$ can be obtained
from the recurrence relation,
\be
\bigg[\hspace{-4pt}\bigg[\left(y_{+}\!\cdot y_{-}-\tfrac12N\,\lambda\right)
\star (y_{+}\!\cdot y_{-})^{n}
\,\bigg]\hspace{-4pt}\bigg]
=\s_{n+1}-\frac N2\,\lambda\,\s_{n}-\frac{n\,(N+n-1)}4\,\s_{n-1}= 0\,.
\label{recur}
\ee
Packing the $\s_{n}$ as \mt{\s(z)=\sum_{n=0}^{\infty} \s_{n}\,z^{n}/n!},
the relation \eqref{recur} becomes a differential equation:
\be\label{diffeqC}
\left[\left(1+\frac{z}2\right)\left(1-\frac{z}2\right)\partial_{z}
-\frac{N}2\left(\frac{z}2+\l\right)\right]\s(z)=0\,,
\ee
whose solution can be easily obtained as
\be
	\s(z)=\left(1-\frac{z}2\right)^{-P}\left(1+\frac{z}2\right)^{-Q}\,,
\ee
with
\be
	P:=N\,\frac{1+\l}2\,,\qquad Q:=N\,\frac{1-\l}2\,.
\ee
However, we need $s(z)$ rather than $\s(z)$\,,
and the former can be obtained from the latter as
\be
	s(z)=(N-1)\int_0^1dw\,(1-w)^{N-2}\,\s(w\,z)\,.
	\label{rel t tau}
\ee
Rewriting $\s(z)$ in an integral form,
\be
\s(z)=\frac{\Gamma(P+Q)}{\Gamma(P)\,\Gamma(Q)}
\int_0^1\,dx\,\frac{x^{P-1}\,(1-x)^{Q-1}}{\big[1+\frac{z}2\,(1-2\,x)\big]^{P+Q}}\,,
\label{int tau}
\ee
and evaluating the $w$-integral first in \eqref{rel t tau}
with \eqref{int tau}, we get
\ba
	s(z)=\frac{\Gamma(N)}{\Gamma(P)\,\Gamma(Q)}
	\int_0^1dx\,\frac{x^{P-1}\,(1-x)^{Q-1}}
	{1+(1-2x)\,\frac{z}2}\,,
\ea
where we used $P+Q=N$\,.

After obtaining the trace of the generating function $\tilde L(\tilde V)$\,,
we can also compute the trace of any Gaussian element
of $hs(\mathfrak{gl}_{N})$\,, that is, $\exp(y_{+}\!\cdot B\cdot y_{-})$
with an arbitrary matrix $B$\,.
Using the identities,
\ba
	&&\exp(y_{-}\!\cdot B\cdot y_{+})
	=g(\partial_{\tilde v_{+}}\!\cdot B\cdot \partial_{\tilde v_{-}})\,
	\exp(y_{-}\!\cdot \tilde V\cdot y_{+})\,\big|_{\tilde v=0}\,,\\
	&&g(\partial_{\tilde v_{+}}\!\cdot B\cdot \partial_{\tilde v_{-}})\,
	\frac1{1-c\,\tilde v_{-}\!\cdot\tilde v_{+}}\,\Big|_{\tilde v=0}
	=\frac1{\det_{N}\left(1-c\,B\right)}\,,
\ea
with $g(z)=\sum_{n=0}^\infty z^{n}/(n!)^{2}$\,, we get
\be
	\tr\big[\exp(y_{-}\cdot B\cdot y_{+})\,\big]
	=\frac{\Gamma(N)}{\Gamma(P)\,\Gamma(Q)}
	\int_0^1dx\,\frac{x^{P-1}\,(1-x)^{Q-1}}
	{\det_{N}\left[1+\tfrac12(1-2x)\,B\right]}\,.
\ee
From this, and using $\star$ product formula for  Gaussian functions,
one can deduce the trace formula for a generic element $f(y)$ in $hs(\mathfrak{gl}_N)$ as
\be
	\tr\big[\,f(y)\,\big]
	=(\D_{\lambda} \star f)(0)\,,
	\label{gen trace sl}
\ee
where $\D_{\lambda}$ is given by
\be
	\D_{\lambda}(y)=
	\frac{\Gamma(N)}{\Gamma\big(\frac {N(1+\l)}2\big)\,
	\Gamma\big(\frac{N(1-\l)}2\big)}
	\int_0^1dx\
	x^{\frac{N(1+\l)}2-1}\,(1-x)^{\frac{N(1-\l)}2-1}\,e^{2(1-2x)\,y_{+}\cdot\,y_{-}}\,.
	\label{D sl}
\ee
Notice that  $\D_{\lambda}$ is nothing but
the deformed version of the $\mathfrak{gl}_{1}$ projector introduced in \cite{Vasiliev:2001wa,Alkalaev:2002rq}.\footnote{See also \cite{Sagnotti:2005ns} for an attempt of a different formulation for the coset algebra.}
Hence, retrospectively, the formula \eqref{gen trace sl} is very natural
extension of the $hs(\mathfrak{sp}_{2N})$ trace \eqref{sp trace}  to $hs_\l(\frak{sl}_N)$\,.

\subsubsection*{Structure constant}

Let us come back to the relation \eqref{prod sl}.
Since the $\lstar$ product can be replaced with the $\star$ product inside of the trace,
for the $n$-linear forms, it is sufficient to compute
\mt{a_{1}\star\cdots\star a_{n}}
where $a_i$ are generating functions of $hs_{\l}(\frak{sl}_{N})$ generators.
The generating function admits again a simple Gaussian form due to \eqref{min para}:
\be
	L(V)=\exp\left(y_{-}\!\cdot V\cdot y_{+}\right).
	\label{L gen}
\ee
Hence, we need to compute
\be
	\frac{1}{G^{\sst(n)}(\rho,V_{1},\ldots,V_{n})}:=
	e^{2\rho\,y_{+}\cdot\,y_{-}}\star L(V_{1})\star \cdots
	\star L(V_{n})\,
	\Big|_{y=0}\,.
	\label{sl n star}
\ee
Here we introduced a $\star$ product of \mt{e^{2\rho\,y_{+}\!\cdot\, y_{-}}}
and evaluated at the end with $y=0$\,,
since this is exactly the necessary information to compute the trace using \eqref{gen trace sl}.
For the evaluation of the $\star$ product
in \eqref{sl n star}, we can simply use the rule of $hs(\mathfrak{sp}_{2N})$
with $U^{\a a\: \b b}=\epsilon^{\a\b}\,V^{a}{}_{c}\,\eta^{cb}$
since $\mathfrak{sl}_{N}$ is a subalgebra of $\frak{sp}_{2N}$\,.
In this way, we obtain
\ba
	G^{\sst(n)}(\rho,V)\eq
	\frac
	{\det_{N}\left(\frac12\,\frac{1+\rho}{1-\rho}\,
	\prod_{k=1}^{n}\frac{2+V_{k}}{2-V_{k}}+\frac12\right)}
	{\det_{N}\left(\frac12\,\frac{1+\rho}{1-\rho}+\frac12\right)
	\prod_{k=1}^{n}\det{}_{\!N}\left(\frac12\,\frac{2+V_{k}}{2-V_{k}}+\frac12\right)} \nn
	\eq \det{}_{\!N}\left[\,
	\frac{1+\rho}2\,\prod_{k=1}^{n}(1+\,V_{k})
	+\frac{1-\rho}2\right],
	\label{sl G}
\ea
where we used the condition $V^2=0$ for the last equality.
Again the evaluation of the determinant for $n=2$ is immediate
and gives
\be
	G^{\sst (2)}(\rho,V)	
	=1+
	\left(\frac{1-\rho}2\right)\left(\frac{1+\rho}2\right)
	\la V_{1}\,V_{2}\ra\,.
\ee
Using the trace formula (\ref{gen trace sl}\,,\,\ref{D sl}),
we end up with the following expression of the bilinear form:
\ba\label{trace2}
\mathcal B(V)
\eq \frac{\Gamma(N)}{\Gamma(P)\,\Gamma(Q)}\int_0^1\,dx\,\frac{x^{P-1}\,(1-x)^{Q-1}}
{1+\,x\,(1-x)\,\la V_{1}\,V_{2}\ra}\nn
\eq {}_3F_2\Big(\tfrac{N}2\,(1+\l)\,,\,\tfrac{N}2\,(1-\l)\,,\,1\,;\,
\tfrac{N}2\,,\,\tfrac{N+1}2\,;\,-\tfrac14\,\la V_{1}\,V_{2}\ra\Big)\,.
\label{BL sl}
\ea
In order to compute the trilinear form, we need to evaluate first
$G^{\sst (3)}(\rho,V)$.
After some computations described in Section \ref{sec: det},
we get
\be
	G^{\sst (3)}(\rho,V)=
	1+\left(\frac{1-\rho}2\right)\left(\frac{1+\rho}2\right)
	\left[\,\frac{1-\rho}2\,\L_{+}(V)+\frac{1+\rho}2\,\L_{-}(V)\,\right],
	\label{3 det sl}
\ee
where $\L_{\pm}(V)$ are defined by
\ba
	&& \L_{+}(V):=\la V_{1}\,V_{2}\ra+\la V_{2}\,V_{3}\ra+\la V_{3}\,V_{1}\ra
	+\la V_{1}\,V_{2}\,V_{3}\ra\,,\nn
	&&
	\L_{-}(V):=\la V_{1}\,V_{2}\ra+\la V_{2}\,V_{3}\ra+\la V_{3}\,V_{1}\ra
	-\la V_{3}\,V_{2}\,V_{1}\ra\,.
\ea
Again, using the trace formula (\ref{gen trace sl}\,,\,\ref{D sl}), we get
the trilinear form as
\be
\mathcal C(V)
=\frac{\Gamma(N)}{\Gamma(P)\,\Gamma(Q)}\int_0^1\,dx\,\frac{x^{P-1}\,(1-x)^{Q-1}}
{1+\,x\,(1-x)\,\big[\,x\,\L_{+}(V)+(1-x)\,\L_{-}(V)\,\big]}\,.
\label{TL sl}
\ee
The integral can be evaluated by expanding the denominator and one gets
\be
\mathcal C(V)=\sum_{k=0}^{\infty}\,\sum_{\ell=0}^{k}\,
(-1)^{k}\,\binom{k}{\ell}\,
\frac{\big(\tfrac{N(1+\l)}2\big)_{2\,k-\ell}
\big(\tfrac{N(1-\l)}2\big)_{k+\ell}}
{(N)_{3k}}\,
\big[\L_{+}(V)\big]^{k-\ell}\,\big[\L_{-}(V)\big]^{\ell}\,,
\ee
which is a double series in $\L_+(V)$ and $\L_-(V)$.
One can see that all the results are symmetric in $\lambda\to -\lambda$
and this shows $hs_{-\lambda}(\mathfrak{sl}_{N})\simeq
hs_{\lambda}(\mathfrak{sl}_{N})$\,.
Incidentally, let us recall that in $5D$\,,
$hs(\nu)=hs_{+\lambda}(\mathfrak{sl}_{4})\oplus
hs_{-\lambda}(\mathfrak{sl}_{4})$
with \eqref{nu lambda}.

\subsection{$hs(\mathfrak{so}_{N})$}
\label{sec: so}

Finally, we consider $hs(\mathfrak{so}_{N})$\,, the most relevant case for physics,
where we need to handle the $\mathfrak{sp}_{2}$ coset.

\subsubsection*{Trace}

We begin again with the general definition \eqref{trace} of the trace.
As in the $hs_{\l}(\mathfrak{sl}_{N})$ case, we first consider
the trace of the generating function $\tilde M(\tilde W)$
\eqref{gen M tilde} of $\widetilde{hs}(\mathfrak{so}_{N})$\,:
\be
	\tr\big[\,\tilde M(\tilde W)\,\big]=\tr\left[\,\exp\!\left(\,y_{-}\!\cdot \tilde W\cdot y_{+}\right)\right]
	=t\left(\la \tilde W^{2}\ra\right),
	\qquad t(z)= \sum_{n=0}^{\infty}\,t^{\sst (2n)}_{n}\,\frac{z^{n}}{n!}\,,
	\label{b z}
\ee
which is given by the function $t(z)$ that has
the Taylor expansion coefficients
$t^{\sst (2n)}_{n}$ appearing in \eqref{rep M}.
By taking the maximal trace of \eqref{rep M},
we get the relation,
\ba
	&&t^{\sst (2n)}_{n}\,
	\big(\,\partial_{\tilde w_{+}}\!\cdot \partial_{\tilde w_{[-}}\,
	\partial_{\tilde w_{+]}}\!\cdot \partial_{\tilde w_{-}}\big)^{n}
	\,\frac{\left(\,\tilde w_{+}\!\cdot \tilde w_{[-}\,\tilde w_{+]}\!\cdot \tilde w_{-}\right)^{n}}{n!}\nn
	&&=\,\bigg[\hspace{-4pt}\bigg[\,
	\big(\,\partial_{\tilde w_{+}}\!\cdot \partial_{\tilde w_{[-}}\,
	\partial_{\tilde w_{+]}}\!\cdot \partial_{\tilde w_{-}}\big)^{n}
	\frac{\left(\,
	y_{+}\!\cdot \tilde w_{[-}\,\tilde w_{+]}\!\cdot y_{-}\right)^{2n}}{(2n)!}
	\,\bigg]\hspace{-4pt}\bigg]\,.
\ea
whose simplification reads
\be
	t^{\sst (2n)}_{n}=\frac{n!}{\left(\frac N2\right)_{n}
	\left(\frac{N-1}2\right)_{n}}\,\t_{n}\,,\qquad
	\t_{n}:=\bigg[\hspace{-4pt}\bigg[
	\left(\frac{y_{+}\!\cdot y_{[-}\,y_{+]}\!\cdot y_{-}}4\right)^{n}
	\bigg]\hspace{-4pt}\bigg]\,.
\ee
The sequence $\t_{n}$
can be determined by setting up a recurrence relation using \eqref{so K} as
\ba
	\bigg[\hspace{-4pt}\bigg[\,
	\frac14\,y_{[+}\!\cdot y_{[-}\,\star\,
	y_{+]}\!\cdot y_{-]}
	\left(\frac{
	y_{+}\!\cdot y_{[-}\,y_{+]}\!\cdot y_{-}}4\right)^{\!n}
	\bigg]\hspace{-4pt}\bigg]=
	\t_{n+1}- \frac18\left(\frac 32+n\right)\left(\frac N2+n\right)\t_{n}=0\,,
	\qquad
\ea
and finally we obtain the coefficients $t^{\sst (2n)}_{n}$ as
\be
	t^{\sst (2n)}_{n}=\frac{n!\,\big(\tfrac 32\big)_{n}}{8^{n}\left(\frac{N-1}2\right)_{n}}\,.
\ee
Coming back to the function $t(z)$ in \eqref{b z},
we get a hypergeometric function which
can be represented by the following integral:
\be
	t(z)
	={}_{2}F_{1}\big(\,1\,,\,\tfrac32\,;\,\tfrac{N-1}2\,;\,\tfrac z8\,\big)
	=
	\frac{\Gamma\!\left(\frac {N-1}2\right)}{\Gamma\!\left(\frac32\right)
	\Gamma\!\left(\frac{N-4}2\right)}\,
	\int_{0}^{1} dx\,\frac{x^{\frac12}\,(1-x)^{\frac{N-6}2}}{1-x\,\tfrac z8}\,.
\ee
Now let us move to the trace of a Gaussian element $\exp(y_{-}\!\cdot C\cdot y_{+})$ of $\widetilde{hs}(\mathfrak{so}_{N})$
given by an arbitrary antisymmetric matrix $C$\,.
Using  the identities,
\ba
	&& \exp\!\left(\,y_{-}\!\cdot C\cdot y_{+}\right)
	=h\left(\partial_{\tilde w_{+}}\!\cdot C\cdot \partial_{\tilde w_{-}}\right)\,
	\exp\!\left(\,y_{+}\!\cdot \tilde w_{[-}\,\tilde w_{+]}\!\cdot y_{-}\right)\Big|_{u=0}\,,\\
	&& h\left(\partial_{\tilde w_{+}}\!\cdot C\cdot \partial_{\tilde w_{-}}\right)\,
	\frac1{1-\frac{c^{2}}2\,\tilde w_{+}\!\cdot \tilde w_{[-}\,\tilde w_{+]}\!\cdot \tilde w_{-}}\,\bigg|_{\tilde w=0}
	=\frac1{\det_{N}\left(1-c\,C\right)}\,,
\ea
with $h(z)=\sum_{n=0}^{\infty} (2\,z)^{n}/[(n+1)!\,n!]$\,,
we obtain its trace as
\be
	\tr\Big[\exp\left(\,y_{-}\!\cdot C\cdot y_{+}\right)\Big]
	=\frac{\Gamma\!\left(\frac {N-1}2\right)}{\Gamma\!\left(\frac32\right)
	\Gamma\!\left(\frac{N-4}2\right)}\,
	\int_{0}^{1} dx\,\frac{x^{\frac12}\,(1-x)^{\frac{N-6}2}}
	{\det_{N}\left(1-\frac{\sqrt{x}}2\,C\right)}\,.
\ee
This formula can be also recast into more intuitive form as
\be
	\tr\big[\,f(y)\,\big]
	=(\D\star f )(0)\,,
	\label{so trace}
\ee
where $\D$ corresponds this time to the $\mathfrak{sp}_{2}$ projector,
\be
	\D(y)=\frac{\Gamma\!\left(\frac {N-1}2\right)}{\Gamma\!\left(\frac32\right)
	\Gamma\!\left(\frac{N-4}2\right)}\,
	\int_{0}^{1} dx\,x^{\frac12}\,(1-x)^{\frac{N-6}2}
	\,e^{-2\sqrt{x}\,y_{+}\cdot\,y_{-}}\,.
	\label{D so}
\ee
Notice however the above expression of the projector $\D$ differs from the original one given in \cite{Vasiliev:2004cm}.
The latter, denoted here by $\tilde \D$\,, has the form,
\be
	\tilde \Delta(y)=\frac{\Gamma\left(\frac{N-1}2\right)}
	{\Gamma\left(\frac12\right)
	\Gamma\left(\frac{N-2}2\right)}\,
	\int^{1}_{-1} ds\,(1-s^{2})^{\frac{N-4}2}\,
	\cos\left(s\,\sqrt{2\,y_{\a}\!\cdot y_{\b}\,
	y^{\a}\!\cdot y^{\b}}\right).
	\label{projec}
\ee
A noticeable difference is that the expression \eqref{D so} does not have a $\mathfrak{sp}_{2}$-invariant form
involving only $y_{+}\!\cdot y_{-}$
--- not
\mt{y_{\a}\!\cdot y_{\b}\ y^{\a}\!\cdot y^{\b}}\,. However, as shown in Appendix \ref{sec: projector}, the two expressions are equivalent
in the sense of
\be
	(\D\star f )(0)=(\tilde \D\star f)(0)\qquad
	\forall f\in \widetilde{hs}(\mathfrak{so}_{N})\,.
\ee
In the following, we shall use the expression \eqref{D so}
as it leads to simpler computations.

\subsubsection*{Structure constant}

In order to obtain the bilinear and trilinear forms,
we need to compute the $\star$ product of
the generating function of HS generators, which admit again a simple form:
\be
	M(W)=\exp\left(y_{-}\!\cdot W\cdot y_{+}\right),
	\label{X gen}
\ee
as a function of the minimal orbit element $W$ \eqref{min orbit}.
For the use of the trace formula (\ref{so trace}\,,\,\ref{D so}),
the Gaussian factor \mt{e^{2\r\,y_{+}\!\cdot\, y_{-}}} should be again inserted in the computation,  hence we consider
\be
	\frac{1}{G^{\sst(n)}(\r,W_{1},\ldots,W_{n})}:=
	e^{2\r\,y_{+}\cdot\,y_{-}}\star M(W_{1})\star \cdots
	\star M(W_{n})\,
	\Big|_{y=0}\,.
\ee
Since $\mathfrak{so}_{N}$ is a subalgebra of $\mathfrak{sl}_{N}$\,,
we can use the formula \eqref{sl G} with the simple replacement
$V^{ab}=W^{ab}$ ending up with
\be
	G^{\sst(n)}(\r,W)
	= \det{}_{\!N}\left[\,
	\frac{1+\r}2\,\prod_{k=1}^{n}(1+W_{k})
	+\frac{1-\r}2\right].
\ee
Again, the $n=2$ case can be evaluated simply as
\be
	G^{\sst (2)}(\r,W)	
	=\left[\,1+\frac{1-\r^{2}}8\,
	\la W_{1}\,W_{2}\ra\,\right]^{2},
\ee
and gives the following expression for the bilinear form:
\ba
	\mathcal B(W)\eq
	\frac{\Gamma\!\left(\frac {N-1}2\right)}{\Gamma\!\left(\frac32\right)
	\Gamma\!\left(\frac{N-4}2\right)}\,
	\int_{0}^{1} dx\,\frac{(1-x)^{\frac12}\,x^{\frac{N-6}2}}{
	\left(1+
	\frac x8\,
	\la W_{1}\,W_{2}\ra\right)^{2}}\nn
	\eq {}_{2}F_{1}\Big(
	2\,,\,\tfrac{N-4}{2}\,;\,\tfrac{N-1}{2}\,;\,-\tfrac18\,\la W_{1}\,W_{2} \ra \Big)\,,
	\label{BL so}
\ea
which has been obtained in \cite{Vasilev:2011xf} in a different notation.
The $n=3$ case requires more involved computations --- see Section \ref{sec: det} ---
and we get in the end
\be
	G^{\sst (3)}(\r,W)	
	= \left[1+\frac{1-\r^{2}}8\,
	\L(W)\right]^{2}-\frac{\r^2\,(1-\r^{2})^{2}}{32}\,\S(W)\,,
	\label{3 det so}
\ee
where $\L(W)$ and $\S(W)$ are given by
\ba
	&&\L(W) := \la W_{1}\,W_{2}\ra+\la W_{2}\,W_{3}\ra
	+\la W_{3}\,W_{1}\ra+\,\la W_{1}\,W_{2}\,W_{3}\ra\,,\\
	&&\S(W):= \left\la (W_{1}\,W_{(2}\,W_{3)})^{2}\right\ra
	=\frac12\,\la W_{1}\,W_{2}\,W_{3}\ra^{2}
	+\frac14\,\la W_{1}\,W_{2}\ra
	\la W_{2}\,W_{3}\ra \la W_{3}\,W_{1}\ra\,.
	\label{Sigma}
\ea
Finally, the trilinear form is given by
\be
	\mathcal C(W)=
	\frac{\Gamma\!\left(\frac {N-1}2\right)}{\Gamma\!\left(\frac32\right)
	\Gamma\!\left(\frac{N-4}2\right)}\,
	\int_{0}^{1} dx\,\frac{(1-x)^{\frac12}\,x^{\frac{N-6}2}}{
	\left(1+
	\frac x8\,\L(W)\right)^{2}
	-\frac{(1-x)\,x^{2}}{32}\,\S(W)}\,,
	\label{TL so}
\ee
and the integral can be evaluated, by expanding the denominator, as
\be
	\mathcal C(W)=\sum_{m=0}^{\infty}\sum_{n=0}^{\infty}
	\frac{\left(\frac{N-4}2\right)_{m+2n}
	(2)_{m+2n}}{\left(\frac{N-1}2\right)_{m+3n}\,8^{m+3n}}\,
	\frac{\left[-\L(W)\right]^{m}}{m!}\,
	\frac{\left[4\,\S(W)\right]^{n}}{n!}\,.
\ee

With this, we have completed
the computations of the bilinear and trilinear forms
of the HS algebras associated with classical Lie algebras.
In the next subsections, we provide an important element
left out
in the previous computations --- the evaluation of determinant ---
and examine certain consistency conditions for our results.

\subsection{Evaluation of determinant}
\label{sec: det}

In the previous sections \ref{sec: sp}, \ref{sec: sl}
and \ref{sec: so}, we faced the evaluation of the determinant,
\be
	G^{\sst(n)}(\rho,A)=\det\left[\,
	\frac{1+\r}2\,\prod_{i=1}^{n}(1+A_{i})
	+\frac{1-\r}2\right],
\ee
for the computations of  the $n$-linear forms.
Here, the matrices $A_{i}$ are in the minimal coadjoint orbit,
so  either $U_{i}$, $V_{i}$ or $W_{i}$ depending on whether we
consider $\frak{sp}_{2N}$\,, $\frak{sl}_{N}$ or $\frak{so}_{N}$\,.
%\footnote{This determinant  can be computed  in various
%different ways.
%One way would be recasting it as a Gaussian integral,
%and performing a kind of perturbative expansion.
%This procedure gives a graphic representations
%of $n$-point functions
%reminiscent of
%discussed in \cite{Vasilev:2011xf}.}
Because the minimal orbit matrices $A_{i}$ admit the parameterizations
\eqref{min para}, they enjoy the following property:
\be
	\la\,A_{i_{1}}\,\cdots\,A_{i_{p}} \ra =
	\la \,\bar A_{i_{1}}\cdots\,\bar A_{i_{p}} \ra\,,
\ee
where $(\bar A_{i})_{jk}=\delta_{ij}\,A_{jk}$\,,
and $A_{ij}$ are given by
 \be
	U_{ij}=\Omega_{AB}\,(u_{i})^{A}\,(u_j)^{B}\,,
	\qquad
	V_{ij}=(v_i)_{+}\!\cdot(v_j)_{-}\,,
	\qquad
	(W_{ij})_{\a\b}=\frac12\,(w_{i})_{\a}\!\cdot(w_{j})_{\b}\,.
\ee
Hence, the \mt{2N\times 2N} or \mt{N\times N} matrices
$A_{i}$ can be
replaced by the $n\times n$ ones $\bar A_{i}$\,.
Notice that only for the $\mathfrak{so}_{N}$ case, the components $A_{ij}$ are again $2\times2$ matrices.
After some computations, one can show that
this determinant can be recast into
\be
	G^{\sst (n)}(\r,A)=
	\det\left[1+\frac{1-\r}2\,\mathsf{Up}(A)+\frac{1+\r}2\,
	\mathsf{Lo}(A)\right],
\ee
in terms of upper and lower triangular matrices
$\mathsf{Up}(A)$ and $\mathsf{Lo}(A)$ with components,
\be
	\big[\mathsf{Up}(A)\big]_{ij}=\delta_{i<j}\,A_{ij}\,,
	\qquad
	\big[\mathsf{Lo}(A)\big]_{ij}=\delta_{i>j}\,A_{ij}\,.
\ee
This expression makes simple the evaluation of the determinant.
Focusing on the $n=3$ case, we get
\ba
	G^{\sst (3)}(\r,A)
	\eq \det\bigg[
	1+\left(\frac{1-\r}2\right)\left(\frac{1+\r}2\right)\,\Big\{
	A_{12}\,A_{21}+A_{23}\,A_{32}+A_{31}\,A_{13}\nn
	&& \hspace{90pt}
	+\,\frac{1-\r}2\,A_{12}\,A_{23}\,A_{31}
	-\frac{1+\r}2\,A_{13}\,A_{32}\,A_{21}\Big\}\,\bigg]\,.
\ea
For $\mathfrak{sp}_{2N}$ and $\mathfrak{sl}_{N}$,
this is the end of the computation, and using
\be
	A_{ij}\,A_{ji}=\la A_{i}\,A_{j} \ra\,,
	\qquad
	A_{ij}\,A_{jk}\,A_{ki}
	=\la A_{i}\,A_{j}\,A_{k}\ra\,,
\ee
we obtain the results \eqref{3 det sp} and \eqref{3 det sl}.
For $\mathfrak{so}_{N}$, one still needs to evaluate
the determinant of  $2\times 2$ matrix,
and a straightforward computation gives \eqref{3 det so}.

\subsection{Isomorphisms between HS algebras}

Let us conclude this section by examining the
HS algebras associated with isomorphic classical Lie algebras.
This can be considered as consistency checks for our results.

\paragraph{\underline{$\bm{\mathfrak{sl}_{2}\simeq \mathfrak{sp}_{2}}$ case}}
To begin with, we consider the case $\mathfrak{sl}_{2}\simeq \mathfrak{sp}_{2}$\,.
The bilinear and trilinear forms of $hs(\mathfrak{sp}_{2})$ are both given by the function $1/\sqrt{1+z}$ but with
different arguments --- see \eqref{sp B C}.
In the case of $\mathfrak{sl}_{N}$, they are given in general by
different functions, but for \mt{N=2}
due to the identity,
\be
	\la V_{1}\,V_{2}\,V_{3}\ra=-\la V_{3}\,V_{2}\,V_{1}\ra\,,
	\label{sl2 prop}
\ee
they do admit expressions through the same function.
In particular, this function coincides to that of $\mathfrak{sp}_{2}$
when the deformation parameter is $\lambda=1/2$\,:
\be
 	{}_3F_2\Big(\tfrac{N}2\,(1+\l)\,,\,\tfrac{N}2\,(1-\l)\,,\,1\,;\,
	\tfrac{N}2\,,\,\tfrac{N+1}2\,;\,-z\Big)
	=\frac1{\sqrt{1+z}}\qquad \big[N=2, \l=\tfrac12\big]\,.
\ee
Hence, this shows the isomorphism
$hs_{\frac12}(\mathfrak{sl}_{2})\simeq hs(\mathfrak{sp}_{2})$\,.

\paragraph{\underline{$\bm{\mathfrak{so}_{5}\simeq \mathfrak{sp}_{4}}$
case}}
Let us move to the cases involving $\mathfrak{so}_{N}$\,.
For that, let us first note that $\S(W)$ in \eqref{Sigma} can be also written as
\ba
	\S(W)
	=-45\,
	(W_{1})^{a_{1}}{}_{[a_{1}}\,
	(W_{2})^{a_{2}}{}_{a_{2}}\,
	(W_{3})^{a_{3}}{}_{a_{3}}\,
	(W_{1})^{a_{4}}{}_{a_{4}}\,
	(W_{2})^{a_{5}}{}_{a_{5}}\,
	(W_{3})^{a_{6}}{}_{a_{6}]}\,,
\ea
therefore vanishes for $N$ smaller than 6.
Consequently, in such cases the bilinear and trilinear forms are both given
by the same function.
In particular, for $N=5$\,, this function coincides with that of $\mathfrak{sp}_{4}$\,:
\be
	{}_{2}F_{1}\Big(
	2\,,\,\tfrac{N-4}{2}\,;\,\tfrac{N-1}{2}\,;\,-z \Big)
	=\frac1{\sqrt{1+z}}\qquad \big[N=5\big]\,.
\ee
This demonstrates the isomorphism $hs(\mathfrak{so}_{5})\simeq hs(\mathfrak{sp}_{4})$\,.
\paragraph{\underline{$\bm{\mathfrak{so}_{6}\simeq \mathfrak{sl}_{4}}$ case}}
The bilinear form and the trilinear form
of $hs(\mathfrak{so}_{6})$ are given by different functions,
and both of them have to coincide with
those of $hs(\mathfrak{sl}_{4})$\,.
In order to check these, we need first to establish the link
$\mathfrak{so}_6\simeq \mathfrak{sl}_4$
using the chiral spinor representation
$\S_{ab}$ as
\be
M_{ab}=-L^{\alpha}{}_{\b}\,(\S_{ab})^{\b}{}_{\a}\,,
\qquad
V_{\a}{}^{\b}=-\frac12\,W_{ab}\,(\S^{ab})^{\b}{}_{\a}\,.
\ee
From this, we get the relation between
the arguments of the bilinear forms (\ref{BL so}\,,\,\ref{BL sl})\,:
\be
	\la\,W_{1}\,W_{2}\,\ra=2\,\la\,V_{1}\,V_{2}\,\ra\,,
\ee
and the function appearing in the bilinear form
of $hs(\mathfrak{so}_{6})$ coincides
with that of $hs_{\l}(\mathfrak{sl}_{4})$ when $\lambda=0$\,:
\ba
	&{}_{2}F_{1}\Big(
	2\,,\,\tfrac{N-4}{2}\,;\,\tfrac{N-1}{2}\,;\,-z\Big)
	={}_3F_2\Big(\tfrac{N'}2\,(1+\l)\,,\,\tfrac{N'}2\,(1-\l)\,,\,1\,;\,
	\tfrac{N'}2\,,\,\tfrac{N'+1}2\,;\,-z\Big)\nn
	&\big[N=6,\, N'=4,\, \l=0\big]\,.
\ea
For the trilinear forms, we have the following relations
between the arguments of the trilinear forms \eqref{TL so} and \eqref{TL sl}\,:
\be
	\L(W)=\L_+(V)+\L_-(V)\,,\qquad
	\S(W)=\frac12\,\big(\L_+(V)-\L_-(V)\big)^2\,,
\ee
and the coincidence of the trilinear forms
can be shown thanks to the identity:
\be
	\int_{0}^{1} dx\,\frac{(1-x)^{\frac12}}{
	\left(1+x\,z\right)^{2}
	-(1-x)\,x^{2}\,\o^{2}} =\int_0^1\,dx\,\frac{4\,x\,(1-x)}{1+4\,x\,(1-x)\,
	\big[\,z+(2x-1)\,\o\,\big]}\,,
\ee
which can be proven by expanding both integrands in $z$ and $\o$
and evaluating the integrals.
This demonstrates the isomorphism $hs(\mathfrak{so}_{6})
\simeq hs_{0}(\mathfrak{sl}_{4})$\,.

\section{More on $hs_{\lambda}(\mathfrak{sl}_{N})$}
\label{sec: more sl}

Differently from $hs(\mathfrak{sp}_{2N})$
and $hs(\mathfrak{so}_{N})$\,,
the HS algebra $hs_{\lambda}(\mathfrak{sl}_{N})$
has one-parameter family.
This gives rise to an interesting consequence:
for certain values of $\lambda$, the algebra develops an ideal,
with a finite-dimensional algebra as the corresponding coset one.
This is what happens to the $3D$ algebra $hs[\lambda]
\simeq hs_{\lambda}(\mathfrak{sl}_{2})$\,,
and to the $5D$ algebra
$hs_{\lambda}(\mathfrak{sl}_{4})$\,. In this section we look into these points more closely.

\subsection{Ideals and finite-dimensional HS algebras}
\label{sec: trun}

From the expression \eqref{trace2} of the bilinear form,
one can notice that the hypergeometric function ${}_{3}F_{2}$ becomes
a polynomial when $N(1\pm\lambda)/2$ takes negative integer values.
In such a case,
\be
	N(1\pm\lambda)=-2\,M \qquad [\,M\in \mathbb N\,]\,,
\ee
the invariant bilinear form becomes degenerate for the generators $L^{\sst (n)}$ with $n> M$ implying that they form an ideal.
This ideal itself can be considered as a HS algebra
although it does not contain the generators
corresponding to the fields of spin $s\le M+1$
--- however, if needed, one can simply include the spin-two generators to this algebra
with standard commutation relations analogous to \eqref{HS commut}.

On the other hand, one can also consider the coset of the original algebra by this ideal. The resulting algebra is then composed of a finite number of generators,
\be
	L^{a_{1}\cdots a_{n}}_{b_{1}\cdots b_{n}}\qquad [\,n=0,1,\ldots,M\,]\,,
\ee
therefore the associated spins are bounded by $M+1$\,.
All these generators can be packed into
the traceful generators \eqref{tilde L},
\be
	\tilde L^{a_{1}\cdots a_{M}}_{b_{1}\cdots b_{M}}\,,
\ee
in terms of which one can easily realize that this algebra is isomorphic to
\be
	\mathfrak{gl}_{\binom{N+M-1}{M}}\,,
\ee
where \mt{\binom{N+M-1}{M}} corresponds simply
to the number of possible
combinations of  symmetrized index $(a_{1},\ldots,a_{M})$\,.
For the \mt{N=4} case of
$\mathfrak{sl}_{4}\simeq \mathfrak{so}_{6}$\,,
the $5D$ finite-dimensional HS algebras
have been obtained  in \cite{Manvelyan:2013oua} making use of
the decomposition of $\mathfrak{sl}_{\binom{M+3}{M}}$
generators into traceless tensors of $\mathfrak{sl}_{4}$\,.

So far, we have only considered complex Lie algebras, but for these finite-dimensional algebras, it would be also interesting to find the
corresponding real form induced by that of  $\mathfrak{sl}_{N}$\,.
We consider here only a particular case $\mathfrak{su}(N_{1},N_{2})$\,, a real form of $\mathfrak{sl}_{N_{1}+N_{2}}$\,, since it is the most relevant in physics:
$\mathfrak{su}(1,1)\simeq \mathfrak{sl}(2,\mathbb R)\simeq \mathfrak{so}(1,2)$
and $\mathfrak{su}(2,2)\simeq \mathfrak{so}(4,2)$\,.
To deal with $\mathfrak{su}(N_{1},N_{2})$\,, we simply divide the indices into two groups $\hat a, \hat b =1,\ldots, N_{1}$
and $\check a, \check b=1, \ldots, N_{2}$\,.
Then, the reality conditions of $\mathfrak{su}(N_{1},N_{2})$ read
\be
	\Big(L^{\hat a}_{\hat b}\Big)^{\dagger}=L^{\hat b}_{\hat a}\,,\qquad
	\Big(L^{\check a}_{\check b}\Big)^{\dagger}=L^{\check b}_{\check a}\,,\qquad
	\Big(L^{\hat a}_{\check b}\Big)^{\dagger}=-L^{\check b}_{\hat a}\,.
\ee
From the above, one can deduce
\be
	\left(\tilde L^{\hat a_{1}\cdots\,\hat a_{k}\check a_{k+1}\cdots\,\check a_{M}}_{\hat b_{1}\cdots\,\hat b_{\ell}\check b_{\ell+1}\cdots\,\check b_{M}}\right)^{\dagger}
	=(-1)^{k+\ell}\,
	\tilde L_{\hat a_{1}\cdots\,\hat a_{k}\check a_{k+1}\cdots\,\check a_{M}}^{\hat b_{1}\cdots\,\hat b_{\ell}\check b_{\ell+1}\cdots\,\check b_{M}}\,.
\ee
This reality condition can be also understood in terms of the oscillators as
\be
	\big(y_{\pm \hat a}\big)^{\dagger}=-y_{\mp\hat a}\,,
	\qquad
	\big(y_{\pm \check a}\big)^{\dagger}=y_{\mp\check a}\,.
\ee
Hence, the real HS algebra associated with
$\mathfrak{su}(N_{1},N_{2})$ is
\be
	\mathfrak{u}\big(\,\mathcal N_{\rm even}\,,
	\,\mathcal N_{\rm odd}\,\big)\,,\footnote{The $\mathfrak{u}(1)$ part of $\mathfrak{u}(\mathcal N_{\rm even},\mathcal N_{\rm odd})$ corresponds to  the center of the algebra.
It might be interpreted as the spin-one generator.}
\ee
where
\be
	\mathcal N_{\rm even/odd}
	=\sum_{0\le\,{\rm even/odd}\, k\,\le M}
	\binom{N_{1}+k-1}{k}\,
	\binom{N_{2}+M-k-1}{M-k}\,.
\ee
For $\mathfrak{su}(1,1)$ we get
\be
	\mathfrak{u}\big(\,\tfrac M2+1\,,\,\tfrac{M}2\,\big)
	\quad  [\,{\rm even}\ M\,]\,,
	\qquad
	\mathfrak{u}\big(\,\tfrac{M+1}2\,,\,\tfrac{M+1}2\,\big)
	\quad [\,{\rm odd}\ M\,]\,,
\ee
and, for $\mathfrak{su}(2,2)$ we get
\ba
	&\mathfrak{u}\Big(
	\tfrac{(M+2)(M^{2}+4\,M+6)}{12}
	\,,\,
	\tfrac{M(M+2)(M+4)}{12}\Big)
	\quad &[\,{\rm even}\ M\,]\,,\nn
	&\mathfrak{u}\Big(\tfrac{(M+1)(M+2)(M+3)}{12}\,,\,
	\tfrac{(M+1)(M+2)(M+3)}{12}\Big)
	\quad &[\,{\rm odd}\ M\,]\,.
\ea
We arrive to these real forms when
we repeat all the constructions of the present paper in the real vector space
starting from $\mathfrak{su}(N_{1},N_{2})$\,.
However, let us note that
there is no reason to give preference to these real forms.

\subsection{Reduced set of oscillators}

In Section \ref{sec: def}, the algebras $hs_{\lambda}(\mathfrak{sl}_{N})$
and $hs(\mathfrak{so}_{N})$ are constructed as cosets
making use of $N$ sets of oscillators
which are subject to certain equivalence relations.
In fact, in the case of $hs_{\lambda}(\mathfrak{sl}_{N})$\,,
there exists yet another description where
all the generators can be given by certain polynomials of \mt{N-1} sets of oscillators \cite{Joseph:1974}. Since these oscillators are not subject to any condition,
it is sufficient to know how $\mathfrak{sl}_{N}$ is represented by them: they are simply given by
\be
	L^{N}{}_{j}= y_{+j}\,,\qquad
	L^{i}{}_{j}=y_{-}{}^{i}\,y_{+j}-\frac{\l}N\,\d^{i}_{j}\,, \qquad
	L^{i}{}_{N}=-\,y_{-}{}^{i}\left(y_{-}{}^{j}\, y_{+j}-\l\right),
\ee
with $i,j=1,\ldots, N-1$\,.
Then, the HS generators are given by all possible $\star$ polynomials
of $\mathfrak{sl}_{N}$ generators in the above representation --- which is again the minimal representation.

\subsection{$N=2$ case: the $3D$ HS algebra}

The $N=2$ case is of particular interest since
$hs_{\lambda}(\mathfrak{sl}_{2})$
coincides with the $3D$ HS algebra
 $hs[\lambda]$\,. The latter algebra, also known as
 $Aq(2;\nu)$ \cite{Vasiliev:1989re}, has been investigated
 by many authors --- see \cite{Boulanger:2013zla,Boulanger:2013naa}
 for recent works. In this subsection, we show how the known
 structures of $hs[\lambda]$ can be derived from the results obtained in this paper for $hs_{\lambda}(\mathfrak{sl}_{N})$\,.

 \paragraph{Lone-star product}
 The associative product of $hs[\lambda]$\,, namely Lone-star product  \cite{Pope:1990kc},
 has been derived in the $\mathcal V_{n}^{s}$ basis, which is proportional in our case to
 \be
 	L_{\underbrace{\st1\,\cdots\,1}_{p}
	\underbrace{\st 2\,\cdots\, 2}_{q}}
	\qquad
	[\,p+q=2(s-1)\,,\ p-q=n\,]\,,
\ee
with
\mt{L_{a_{1}\cdots a_{2n}}:=L^{b_{1}\cdots b_{n}}_{a_{1}\cdots a_{n}}\,
	\e_{b_{1}a_{n+1}}\cdots\,\e_{b_{n}a_{2n}}.}
The precise expression for such product is quite lengthy
and we refer to \cite{Pope:1989sr,Pope:1990kc,Fradkin:1990ir}.
Instead, we show how this  $\lstar$ product
can be obtained in a relatively simple form from the bilinear and trilinear forms (\ref{BL sl}\,,\,\ref{TL sl}).
In the $N=2$ case, such forms are simplified
using \eqref{sl2 prop} into\footnote{Moreover, the hypergeometric function admits another simple
expression,
\be
	{}_2F_1\big(\,1+\l\,,\,1-\l\,;\,\tfrac32\,;\,-\,z\,\big)	
	=\frac{\sinh\left(2\,\l\,\sinh^{-1}(\sqrt{z})\right)}{2\,\l\,\sqrt{1+z}\,\sqrt{z}}=
	\frac{\left(\sqrt{1+z}+\sqrt{z}\right)^{2\l}
	-\left(\sqrt{1+z}-\sqrt{z}\right)^{2\l}}{4\,\l\,\sqrt{1+z}\,\sqrt{z}}\,.
\ee
}
\ba\label{Killing3d}
\mathcal B(V)
\eq {}_2F_1\Big(1+\l\,,\,1-\l\,;\,\tfrac32\,;\,-\tfrac14\,\la V_{1}\,V_{2}\ra\Big)\,,\\
\mathcal C(V)
\eq {}_2F_1\Big(1+\l\,,\,1-\l\,;\,\tfrac32\,;\,-\tfrac14\,\L(V)\Big)\,.
\label{BC sl2}
\ea
Let us mention that the bilinear form \eqref{Killing3d} has been initially obtained in \cite{Vasiliev:1989re}
making use of the deformed oscillators.
From \eqref{Killing3d} and \eqref{BC sl2}, we can derive the explicit expression of the $\lstar$ product
for the generating elements:
\be
	L(V_{1})\lstar
	L(V_{2}) =
	\sum_{n=0}^{\infty}
	{}_2F_1\big(\,n+1+\l\,,\,n+1-\l\,;\,n+\tfrac32\,;\,-\tfrac 14\,
	\la V_{1}\,V_{2}\ra\,\big)\,
	L^{\sst(n)}(V_{1}+V_{2}+V_{1}\,V_{2})\,.
\ee
From this, one can extract
the contribution of each generators to the $\lstar$ product.

\paragraph{Deformed oscillators}

Another convenient description of  $hs[\lambda]$
is the deformed oscillators --- see e.g. \cite{Vasiliev:1989re}.
In the following, we provide a link of
 the description presented in Section \ref{sec: def}
 to the deformed oscillator one.
Let us first notice that the former description $hs_{\lambda}(\mathfrak{sl}_{N})$
requires $N$ sets of oscillators $(y_{+a}, y_{-a})$ ---  two sets for $N=2$\,. On the other hand, in the latter description
through deformed oscillators, one needs only one pair of oscillators.
Despite of this discrepancy, one can establish an explicit link between
two descriptions by introducing a single set \emph{matrix-valued} oscillators
out of two sets of usual oscillators as
\be
	\hat{y}_{a}:=2\,\begin{pmatrix} 0 & y_{+a} \\ y_{-}{}^{c}\,\epsilon_{ca} & 0
	\end{pmatrix}.
\ee
Then, we can define the product between two such oscillators as
the matrix product,
\be
	\hat{y}_{a}\ \hat{y}_{b}:=2
	\begin{pmatrix} 0 & y_{+a} \\ y_{-}{}^{c}\,\e_{ca} & 0
	\end{pmatrix}
	\lstar
	2
	\begin{pmatrix} 0 & y_{+b} \\ y_{-}{}^{d}\,\e_{db} & 0
	\end{pmatrix},
\ee
where the multiplications of matrix entities are with respect to the $\lstar$ product.
The commutator of such product is readily calculated and gives
\be
	\big[\,\hat{y}_{a}\,,\,\hat{y}_{b}
	\,\big]=2\,\epsilon_{ab}\left(\,\hat 1 + \hat \nu\,\hat{k}\,\right),
	\label{deform1}
\ee
where $\hat k$ and $\hat \n$ are defined by
\be
	\hat{k}:=\begin{pmatrix} 1 & \,0 \\ 0 & -1
	\end{pmatrix},
	\qquad
	\hat \nu:=2\,\l\, \hat 1+\hat k\,,
\ee
so satisfy
\be
	\hat k^{2}=1\,,\qquad \big\{\, \hat k\,,\,\hat y_{a}\,\big\}=0\,,
	\qquad
	\big[\,\hat{\nu}\,,\,\hat{y}_{a}
	\,\big]=0\,,
	\qquad
	\big[\,\hat{\nu}\,,\,\hat{k}
	\,\big]=0\,.
	\label{deform2}
\ee
Notice that the equations \eqref{deform1} and \eqref{deform2} are
the defining relations of the deformed oscillators, and one can
see that $\hat \n$ is a constant diagonal matrix:
\be
	\hat \nu
	=
	\begin{pmatrix}
	2\,\lambda+1 & 0 \\ 0 & 2\,\lambda-1
	\end{pmatrix},
\ee
and can be treated as a constant number $\hat \n=2\,\lambda \pm 1$
in the $\pm$ eigenspace of $\hat k$\,.

\section{Outlook}
\label{sec: outlook}

Finally, let us conclude the present paper with a few remarks:
\begin{itemize}
\item
First of all, the results obtained here may be applied to
the construction of
 $n$-point correlation functions of the Vasiliev theory,
along the line of \cite{Colombo:2012jx,Didenko:2012tv,Didenko:2013bj}:
the necessary ingredients there are the trace formula and the boundary-to-bulk propagator in the unfolded formulation. The former is provided in this paper, while the latter
has been  investigated in \cite{Didenko:2012vh}.
\item
In this paper, we considered the HS algebras  associated only with symmetric tensor fields.
However in higher dimensions, there exist many other massless fields
of the mixed-symmetry tensor type, and
their understanding is important for the generalization of  the currently understood version of HS theory to wider context.
In this respect, it would be interesting
to generalize our results to the cases of mixed-symmetry HS algebras.
In particular, examining possibilities to interpret the works \cite{Vasiliev:2004cm} and \cite{Boulanger:2011se}
within this picture would be tempting.
\item
Another direction of generalizing HS gauge theory and
the corresponding global symmetry
is the study of partially-massless HS fields.
Actually these have been already explored by a number of authors:
see \cite{Eastwood:2006,Gover:2009,Michel:2011} for the mathematics literature and \cite{Bekaert:2013zya} for the physics one.
Again, it would be interesting to reformulate such results
in the language presented in this paper, in particular by making use
of a certain reductive dual pair correspondence.

\end{itemize}

We are currently investigating the latter two issues,
and hope to report on them in the near future.

\acknowledgments{
We thank
N. Boulanger, A. Campoleoni, R. Manvelyan, E. Skvortsov, M. Taronna
and S. Theisen
for useful discussions.
We also acknowledge the GGI, Florence workshop on ``Higher spins, Strings and Duality'' where the present work was developed.
The work of K.M. was supported in part by Scuola Normale Superiore,
INFN (I.S. TV12),
the MIUR-PRIN contract 2009-KHZKRX, and the ERC Advanced Investigator Grants no. 226455  ``Supersymmetry, Quantum Gravity and Gauge Fields'' (SUPERFIELDS).
}

\appendix

\section{Two representations of the trace projector}
\label{sec: projector}

In \cite{Vasiliev:2004cm}, the $\mathfrak{sp}_{2}$ projector has been determined
in the form of \eqref{projec}.
In order to see its equivalence to the other form
\eqref{D so}, we begin with
the following transformation:
\be
	\cos\left(s\,\sqrt{2\,K_{\a\b}\,K^{\a\b}}\right)
	=T\left[e^{-2\,\o^{2}\,K_{\a\b}\,K^{\a\b}}\right](s)\,.
\ee
where the linear map $T$ is defined by
\be
	T\left[\o^{2n}\right](s)=\frac{n!}{(2n)!}\,s^{2n}\,.
\ee
Then, using the following identity:
\be
	e^{-2\,\o^{2}\,K_{\a\b}\,K^{\a\b}}
	=\int \frac{d^{3}\vec z}{(2\pi)^{3/2}}\,
	e^{-\frac12\,\vec z^{\,2}}
	\cosh(\sqrt{2}\,\o\, z^{\a\b}\,K_{\a\b})\,,
	\label{K 3}
\ee
with $z^{\pm\pm}=\pm z_{1}+i\,z_{2}$\,,
 and $z^{\pm \mp}=z_{3}$\,,
we get $\cosh$ function with $K_{\a\b}$-linear argument.
It can be shown by a straightforward computation
that
\be
	\left(\cosh(\sqrt{2}\, \o\,z^{\a\b}\,K_{\a\b})\star f \right)(0)
	=\Big(\cosh(2\sqrt{2}\,\o\,|\vec z\,|\,y_{+}\!\cdot y_{-})\star f\Big)(0)\,,
\ee
which allows us to replace the $\mathfrak{sp}_{2}$
element $K_{\a\b}$ by $y_{+}\!\cdot y_{-}$
in \eqref{K 3}\,.
The next steps are the $z^{\a\b}$-integral and the $T$-transformation:
first, the $z^{\a\b}$-integral gives
\be
	\int \frac{d^{3}\vec z}{(2\pi)^{3/2}}\,
	e^{-\frac12\,\vec z^{\,2}}\,
	\cosh(2\sqrt{2}\,\o\,|\vec z\,|\,y_{+}\!\cdot y_{-})
	=\sum_{n=0}^{\infty}\frac{(2n+1)!}{n!}\,\o^{2n}\,
	\frac{(2\,y_{+}\!\cdot y_{-})^{2n}}{(2n)!}\,,
\ee
whose $T$-transformation reads
\be
	\sum_{n=0}^{\infty}(2n+1)\,s^{2n}\,
	\frac{(2\,y_{+}\!\cdot y_{-})^{2n}}{(2n)!}\,.
\ee
Finally, evaluating the $s$-integral \eqref{projec}, we get
\be
	(\tilde \D\star f)(0)
	=\sum_{n=0}^{\infty}\frac{\left(\frac32\right)_{n}}
	{\left(\frac{N-1}2\right)_{n}} \left(\frac{(2\,y_{+}\cdot y_{-})^{2n}}{(2n)!}\star f\right)(0)\,.
\ee
On the other hand, it is straightforward to
obtain the above expression starting from $\Delta$ \eqref{D so}\,:
first we expand $\Delta$ in \mt{y_{+}\!\cdot y_{-}}
(again only even powers contribute to the trace) and
then evaluate the $x$-integral of \eqref{D so}.

\bibliographystyle{JHEP}
\bibliography{ref}

\providecommand{\href}[2]{#2}\begingroup\raggedright\begin{thebibliography}{10}

\bibitem{Vasiliev:1990en}
M.~A. Vasiliev, {\it {Consistent equation for interacting gauge fields of all
  spins in (3+1)-dimensions}},  {\em Phys. Lett.} {\bf B243} (1990) 378--382.

\bibitem{Vasiliev:2003ev}
M.~Vasiliev, {\it {Nonlinear equations for symmetric massless higher spin
  fields in $(A)dS_d$}},  {\em Phys. Lett.} {\bf B567} (2003) 139--151,
  [\href{http://xxx.lanl.gov/abs/hep-th/0304049}{{\tt hep-th/0304049}}].

\bibitem{Vasiliev:2004qz}
M.~Vasiliev, {\it {Higher spin gauge theories in various dimensions}},  {\em
  Fortsch. Phys.} {\bf 52} (2004) 702--717,
  [\href{http://xxx.lanl.gov/abs/hep-th/0401177}{{\tt hep-th/0401177}}].

\bibitem{Gaberdiel:2012uj}
M.~R. Gaberdiel and R.~Gopakumar, {\it {Minimal Model Holography}},  {\em J.
  Phys.} {\bf A46} (2013) 214002,
  [\href{http://xxx.lanl.gov/abs/1207.6697}{{\tt arXiv:1207.6697}}].

\bibitem{Fradkin:1986ka}
E.~Fradkin and M.~A. Vasiliev, {\it {Candidate to the Role of Higher Spin
  Symmetry}},  {\em Annals Phys.} {\bf 177} (1987) 63.

\bibitem{Fradkin:1987ks}
E.~Fradkin and M.~A. Vasiliev, {\it {On the Gravitational Interaction of
  Massless Higher Spin Fields}},  {\em Phys. Lett.} {\bf B189} (1987) 89--95.

\bibitem{Sezgin:2001zs}
E.~Sezgin and P.~Sundell, {\it {Doubletons and 5-D higher spin gauge theory}},
  {\em JHEP} {\bf 0109} (2001) 036,
  [\href{http://xxx.lanl.gov/abs/hep-th/0105001}{{\tt hep-th/0105001}}].

\bibitem{Vasiliev:2001wa}
M.~A. Vasiliev, {\it {Cubic interactions of bosonic higher spin gauge fields in
  $AdS_5$}},  {\em Nucl. Phys.} {\bf B616} (2001) 106--162,
  [\href{http://xxx.lanl.gov/abs/hep-th/0106200}{{\tt hep-th/0106200}}].

\bibitem{Sezgin:2001ij}
E.~Sezgin and P.~Sundell, {\it {7-D bosonic higher spin theory: Symmetry
  algebra and linearized constraints}},  {\em Nucl. Phys.} {\bf B634} (2002)
  120--140, [\href{http://xxx.lanl.gov/abs/hep-th/0112100}{{\tt
  hep-th/0112100}}].

\bibitem{Mikhailov:2002bp}
A.~Mikhailov, {\it {Notes on higher spin symmetries}},
  \href{http://xxx.lanl.gov/abs/hep-th/0201019}{{\tt hep-th/0201019}}.

\bibitem{Eastwood:2002su}
M.~G. Eastwood, {\it {Higher symmetries of the Laplacian}},  {\em Annals Math.}
  {\bf 161} (2005) 1645--1665,
  [\href{http://xxx.lanl.gov/abs/hep-th/0206233}{{\tt hep-th/0206233}}].

\bibitem{Bergshoeff:1989ns}
E.~Bergshoeff, M.~Blencowe, and K.~Stelle, {\it {Area Preserving
  Diffeomorphisms and Higher Spin Algebra}},  {\em Commun. Math. Phys.} {\bf
  128} (1990) 213.

\bibitem{Bordemann:1989zi}
M.~Bordemann, J.~Hoppe, and P.~Schaller, {\it Infinite dimensional matrix
  algebras},  {\em Phys. Lett.} {\bf B232} (1989) 199.

\bibitem{Vasiliev:1989re}
M.~A. Vasiliev, {\it {Higher Spin Algebras and Quantization on the Sphere and
  Hyperboloid}},  {\em Int. J. Mod. Phys.} {\bf A6} (1991) 1115--1135.

\bibitem{Pope:1989sr}
C.~Pope, L.~Romans, and X.~Shen, {\it {$W$(infinity) and the Racah-wigner
  Algebra}},  {\em Nucl. Phys.} {\bf B339} (1990) 191--221.

\bibitem{Pope:1990kc}
C.~Pope, L.~Romans, and X.~Shen, {\it {A New Higher Spin Algebra and the Lone
  Star Product}},  {\em Phys. Lett.} {\bf B242} (1990) 401--406.

\bibitem{Fradkin:1990ir}
E.~S. Fradkin and V.~{\relax Ya}. Linetsky, {\it {Infinite dimensional
  generalizations of simple Lie algebras}},  {\em Mod. Phys. Lett.} {\bf A5}
  (1990) 1967--1977.

\bibitem{Prokushkin:1998vn}
S.~Prokushkin and M.~A. Vasiliev, {\it {3-d higher spin gauge theories with
  matter}},  \href{http://xxx.lanl.gov/abs/hep-th/9812242}{{\tt
  hep-th/9812242}}.

\bibitem{Prokushkin:1998bq}
S.~Prokushkin and M.~A. Vasiliev, {\it {Higher spin gauge interactions for
  massive matter fields in 3-D AdS space-time}},  {\em Nucl. Phys.} {\bf B545}
  (1999) 385, [\href{http://xxx.lanl.gov/abs/hep-th/9806236}{{\tt
  hep-th/9806236}}].

\bibitem{Henneaux:2010xg}
M.~Henneaux and S.-J. Rey, {\it {Nonlinear $W_{infinity}$ as Asymptotic
  Symmetry of Three-Dimensional Higher Spin Anti-de Sitter Gravity}},  {\em
  JHEP} {\bf 1012} (2010) 007, [\href{http://xxx.lanl.gov/abs/1008.4579}{{\tt
  arXiv:1008.4579}}].

\bibitem{Campoleoni:2010zq}
A.~Campoleoni, S.~Fredenhagen, S.~Pfenninger, and S.~Theisen, {\it {Asymptotic
  symmetries of three-dimensional gravity coupled to higher-spin fields}},
  {\em JHEP} {\bf 1011} (2010) 007,
  [\href{http://xxx.lanl.gov/abs/1008.4744}{{\tt arXiv:1008.4744}}].

\bibitem{Gaberdiel:2011wb}
M.~R. Gaberdiel and T.~Hartman, {\it {Symmetries of Holographic Minimal
  Models}},  {\em JHEP} {\bf 1105} (2011) 031,
  [\href{http://xxx.lanl.gov/abs/1101.2910}{{\tt arXiv:1101.2910}}].

\bibitem{Campoleoni:2011hg}
A.~Campoleoni, S.~Fredenhagen, and S.~Pfenninger, {\it {Asymptotic W-symmetries
  in three-dimensional higher-spin gauge theories}},  {\em JHEP} {\bf 1109}
  (2011) 113, [\href{http://xxx.lanl.gov/abs/1107.0290}{{\tt
  arXiv:1107.0290}}].

\bibitem{Iazeolla:2008ix}
C.~Iazeolla and P.~Sundell, {\it {A Fiber Approach to Harmonic Analysis of
  Unfolded Higher-Spin Field Equations}},  {\em JHEP} {\bf 10} (2008) 022,
  [\href{http://xxx.lanl.gov/abs/0806.1942}{{\tt arXiv:0806.1942}}].

\bibitem{Boulanger:2011se}
N.~Boulanger and E.~Skvortsov, {\it {Higher-spin algebras and cubic
  interactions for simple mixed-symmetry fields in AdS spacetime}},  {\em JHEP}
  {\bf 1109} (2011) 063, [\href{http://xxx.lanl.gov/abs/1107.5028}{{\tt
  arXiv:1107.5028}}].

\bibitem{Fradkin:1989yd}
E.~S. Fradkin and V.~{\relax Ya}. Linetsky, {\it {Conformal superalgebras of
  higher spins}},  {\em Ann. Phys.} {\bf 198} (1990) 252--292.

\bibitem{Bekaert:2007mi}
X.~Bekaert, {\it {Higher spin algebras as higher symmetries}},
  \href{http://xxx.lanl.gov/abs/0704.0898}{{\tt arXiv:0704.0898}}.

\bibitem{Bekaert:2008sa}
X.~Bekaert, {\it {Comments on higher-spin symmetries}},  {\em Int. J. Geom.
  Meth. Mod. Phys.} {\bf 6} (2009) 285--342,
  [\href{http://xxx.lanl.gov/abs/0807.4223}{{\tt arXiv:0807.4223}}].

\bibitem{Bekaert:2011js}
X.~Bekaert, {\it {Singletons and their maximal symmetry algebras}},
  \href{http://xxx.lanl.gov/abs/1111.4554}{{\tt arXiv:1111.4554}}.

\bibitem{Maldacena:2011jn}
J.~Maldacena and A.~Zhiboedov, {\it {Constraining Conformal Field Theories with
  A Higher Spin Symmetry}},  {\em J. Phys.} {\bf A46} (2013) 214011,
  [\href{http://xxx.lanl.gov/abs/1112.1016}{{\tt arXiv:1112.1016}}].

\bibitem{Boulanger:2013zza}
N.~Boulanger, D.~Ponomarev, E.~Skvortsov, and M.~Taronna, {\it {On the
  uniqueness of higher-spin symmetries in AdS and CFT}},  {\em Int. J. Mod.
  Phys.} {\bf A28} (2013) 1350162,
  [\href{http://xxx.lanl.gov/abs/1305.5180}{{\tt arXiv:1305.5180}}].

\bibitem{Alba:2013yda}
V.~Alba and K.~Diab, {\it {Constraining conformal field theories with a higher
  spin symmetry in d=4}},  \href{http://xxx.lanl.gov/abs/1307.8092}{{\tt
  arXiv:1307.8092}}.

\bibitem{Stanev:2013qra}
Y.~S. Stanev, {\it {Constraining conformal field theory with higher spin
  symmetry in four dimensions}},  {\em Nucl. Phys.} {\bf B876} (2013) 651--666,
  [\href{http://xxx.lanl.gov/abs/1307.5209}{{\tt arXiv:1307.5209}}].

\bibitem{Eastwood:2006}
M.~{Eastwood} and T.~{Leistner}, {\it {Higher Symmetries of the Square of the
  Laplacian}},  \href{http://xxx.lanl.gov/abs/math/0610610}{{\tt
  math/0610610}}.

\bibitem{Gover:2009}
A.~R. {Gover} and J.~{Silhan}, {\it {Higher symmetries of the conformal powers
  of the Laplacian on conformally flat manifolds}},
  \href{http://xxx.lanl.gov/abs/0911.5265}{{\tt arXiv:0911.5265}}.

\bibitem{Michel:2011}
J.~P. {Michel}, {\it {Higher Symmetries of the Laplacian via Quantization}},
  \href{http://xxx.lanl.gov/abs/1107.5840}{{\tt arXiv:1107.5840}}.

\bibitem{Bekaert:2013zya}
X.~Bekaert and M.~Grigoriev, {\it {Higher order singletons, partially massless
  fields and their boundary values in the ambient approach}},  {\em Nucl.
  Phys.} {\bf B876} (2013) 667--714,
  [\href{http://xxx.lanl.gov/abs/1305.0162}{{\tt arXiv:1305.0162}}].

\bibitem{Vasiliev:2012tv}
M.~Vasiliev, {\it {Multiparticle extension of the higher-spin algebra}},  {\em
  Class. Quant.Grav.} {\bf 30} (2013) 104006,
  [\href{http://xxx.lanl.gov/abs/1212.6071}{{\tt arXiv:1212.6071}}].

\bibitem{Gelfond:2013xt}
O.~Gelfond and M.~Vasiliev, {\it {Operator algebra of free conformal currents
  via twistors}},  {\em Nucl. Phys.} {\bf B876} (2013) 871--917,
  [\href{http://xxx.lanl.gov/abs/1301.3123}{{\tt arXiv:1301.3123}}].

\bibitem{Joseph:1974}
A.~Joseph, {\it {Minimal Realizations and Spectrum Generating Algebras}},  {\em
  Commun. Math. Phys.} {\bf 36} (1974) 325.

\bibitem{Joseph:1976}
A.~Joseph, {\it {The minimal orbit in a simple Lie algebra and its associated
  maximal ideal}},  {\em Ann. scient. \'Ecole Norm. Sup. 4e s\'erie} {\bf 9}
  (1976) 1.

\bibitem{Garfinkle:1982}
D.~Garfinkle, {\em {A new construction of the Joseph ideal}}.
\newblock {PhD Thesis}, {Massachusetts Institute of Technology}, 1982.

\bibitem{Levasseur:1988}
T.~Levasseur and S.~P. Simth, {\it {The Minimal Nilpotent Orbit, the Joseph
  Ideal, and Differential Operators}},  {\em J. Algebra} {\bf 116} (1988) 480.

\bibitem{Binegar:1991}
B.~Binegar and R.~Zierau, {\it {Unitarization of a Singular Representation of
  $SO(p,q)$}},  {\em Commun. Math. Phys.} {\bf 138} (1991) 245.

\bibitem{Braverman:1998}
A.~Bradverman and A.~Joseph, {\it {The Minimal Realization from Deformation
  Theory}},  {\em J. Algebra} {\bf 205} (1998) 13.

\bibitem{Li:2000}
J.-S. Li, {\it {Representation Theory of Lie Groups (ch. Minimal
  representations \& reductive dual pairs)}},  {\em IAS/Park City Math. Ser.}
  {\bf 8} (2000) 291.

\bibitem{Gan:2004}
W.~Gan and G.~Savin, {\it {Uniqueness of the Joseph ideal}},  {\em Math. Res.
  Lett.} {\bf 11} (2004) 589.

\bibitem{Eastwood:2007}
M.~Eastwood, P.~Somberg, and V.~Soucek, {\it Special tensors in the deformation
  theory of quadratic algebras for the classical lie algebras},  {\em J. Geom.
  Phys.} {\bf 57} (2007) 2539.

\bibitem{Fronsdal:2009}
C.~Fronsdal, {\it Deformation quantization on the closure of minimal coadjoint
  orbits},  {\em Lett. Math. Phys.} {\bf 88} (2009) 271.

\bibitem{Todorov:2010md}
I.~Todorov, {\it {Minimal representations and reductive dual pairs in conformal
  field theory}},  {\em AIP Conf. Proc.} {\bf 1243} (2010) 13--30,
  [\href{http://xxx.lanl.gov/abs/1006.1981}{{\tt arXiv:1006.1981}}].

\bibitem{Kobayashi:2013ao}
T.~Kobayashi, {\it Special functions in minimal representations},
  \href{http://xxx.lanl.gov/abs/1301.5505}{{\tt arXiv:1301.5505}}.

\bibitem{Gunaydin:1989um}
M.~Gunaydin, {\it {Singleton and Doubleton Supermultiplets of Space-time
  Supergroups and Infinite spin superalgebras}},  {\em CERN-TH-5500/89} (1989).

\bibitem{Fernando:2009fq}
S.~Fernando and M.~Gunaydin, {\it {Minimal unitary representation of SU(2,2)
  and its deformations as massless conformal fields and their supersymmetric
  extensions}},  {\em J. Math. Phys.} {\bf 51} (2010) 082301,
  [\href{http://xxx.lanl.gov/abs/0908.3624}{{\tt arXiv:0908.3624}}].

\bibitem{Fernando:2010dp}
S.~Fernando and M.~Gunaydin, {\it {Minimal unitary representation of SO*(8) =
  SO(6,2) and its SU(2) deformations as massless 6D conformal fields and their
  supersymmetric extensions}},  {\em Nucl. Phys.} {\bf B841} (2010) 339--387,
  [\href{http://xxx.lanl.gov/abs/1005.3580}{{\tt arXiv:1005.3580}}].

\bibitem{Govil:2013uta}
K.~Govil and M.~Gunaydin, {\it {Deformed Twistors and Higher Spin Conformal
  (Super-)Algebras in Four Dimensions}},
  \href{http://xxx.lanl.gov/abs/1312.2907}{{\tt arXiv:1312.2907}}.

\bibitem{Govil:2014uwa}
K.~Govil and M.~Gunaydin, {\it {Deformed Twistors and Higher Spin Conformal
  (Super-)Algebras in Six Dimensions}},
  \href{http://xxx.lanl.gov/abs/1401.6930}{{\tt arXiv:1401.6930}}.

\bibitem{Alkalaev:2002rq}
K.~Alkalaev and M.~Vasiliev, {\it {N=1 supersymmetric theory of higher spin
  gauge fields in AdS(5) at the cubic level}},  {\em Nucl. Phys.} {\bf B655}
  (2003) 57--92, [\href{http://xxx.lanl.gov/abs/hep-th/0206068}{{\tt
  hep-th/0206068}}].

\bibitem{Vasiliev:2004cm}
M.~Vasiliev, {\it {Higher spin superalgebras in any dimension and their
  representations}},  {\em JHEP} {\bf 0412} (2004) 046,
  [\href{http://xxx.lanl.gov/abs/hep-th/0404124}{{\tt hep-th/0404124}}].

\bibitem{Fronsdal:1978rb}
C.~Fronsdal, {\it {Massless Fields with Integer Spin}},  {\em Phys. Rev.} {\bf
  D18} (1978) 3624.

\bibitem{Fronsdal:1978vb}
C.~Fronsdal, {\it {Singletons and Massless, Integral Spin Fields on de Sitter
  Space (Elementary Particles in a Curved Space. 7.}},  {\em Phys. Rev.} {\bf
  D20} (1979) 848--856.

\bibitem{Bekaert:2005ka}
X.~Bekaert and N.~Boulanger, {\it {Gauge invariants and Killing tensors in
  higher-spin gauge theories}},  {\em Nucl. Phys.} {\bf B722} (2005) 225--248,
  [\href{http://xxx.lanl.gov/abs/hep-th/0505068}{{\tt hep-th/0505068}}].

\bibitem{Bekaert:2006us}
X.~Bekaert, N.~Boulanger, S.~Cnockaert, and S.~Leclercq, {\it {On killing
  tensors and cubic vertices in higher-spin gauge theories}},  {\em Fortsch.
  Phys.} {\bf 54} (2006) 282--290,
  [\href{http://xxx.lanl.gov/abs/hep-th/0602092}{{\tt hep-th/0602092}}].

\bibitem{Joung:2013nma}
E.~Joung and M.~Taronna, {\it {Cubic-interaction-induced deformations of
  higher-spin symmetries}},  \href{http://xxx.lanl.gov/abs/1311.0242}{{\tt
  arXiv:1311.0242}}.

\bibitem{Konstein:1988yg}
S.~Konstein and M.~A. Vasiliev, {\it {Massless Representations and
  Admissibility Condition for Higher Spin Superalgebras}},  {\em Nucl. Phys.}
  {\bf B312} (1989) 402.

\bibitem{Konstein:1989ij}
S.~Konstein and M.~A. Vasiliev, {\it {Extended Higher Spin Superalgebras and
  Their Massless Representations}},  {\em Nucl. Phys.} {\bf B331} (1990)
  475--499.

\bibitem{Manvelyan:2013oua}
R.~Manvelyan, K.~Mkrtchyan, R.~Mkrtchyan, and S.~Theisen, {\it {On Higher Spin
  Symmetries in $AdS_{5}$}},  {\em JHEP} {\bf 1310} (2013) 185,
  [\href{http://xxx.lanl.gov/abs/1304.7988}{{\tt arXiv:1304.7988}}].

\bibitem{GK}
I.~M. Gelfand and A.~A. Kirillov, {\it {Sur les corps li\'es aux alg\`ebres
  enveloppantes des alg\`ebres de Lie}},  {\em Publ. Math. IHES.} {\bf 31}
  (1966) 5--19.

\bibitem{Flato:1978qz}
M.~Flato and C.~Fronsdal, {\it {One Massless Particle Equals Two Dirac
  Singletons: Elementary Particles in a Curved Space. 6.}},  {\em Lett. Math.
  Phys.} {\bf 2} (1978) 421--426.

\bibitem{Vasilev:2011xf}
M.~Vasiliev, {\it {Cubic Vertices for Symmetric Higher-Spin Gauge Fields in
  $(A)dS_d$}},  {\em Nucl. Phys.} {\bf B862} (2012) 341--408,
  [\href{http://xxx.lanl.gov/abs/1108.5921}{{\tt arXiv:1108.5921}}].

\bibitem{Didenko:2003aa}
V.~Didenko and M.~Vasiliev, {\it {Free field dynamics in the generalized AdS
  (super)space}},  {\em J. Math. Phys.} {\bf 45} (2004) 197--215,
  [\href{http://xxx.lanl.gov/abs/hep-th/0301054}{{\tt hep-th/0301054}}].

\bibitem{Sagnotti:2005ns}
A.~Sagnotti, E.~Sezgin, and P.~Sundell, {\it {On higher spins with a strong
  Sp(2,R) condition}},  \href{http://xxx.lanl.gov/abs/hep-th/0501156}{{\tt
  hep-th/0501156}}.

\bibitem{Boulanger:2013zla}
N.~Boulanger, P.~Sundell, and M.~Valenzuela, {\it {A Higher-Spin Chern-Simons
  Theory of Anyons}},  \href{http://xxx.lanl.gov/abs/1311.4589}{{\tt
  arXiv:1311.4589}}.

\bibitem{Boulanger:2013naa}
N.~Boulanger, P.~Sundell, and M.~Valenzuela, {\it {Three-dimensional
  fractional-spin gravity}},  \href{http://xxx.lanl.gov/abs/1312.5700}{{\tt
  arXiv:1312.5700}}.

\bibitem{Colombo:2012jx}
N.~Colombo and P.~Sundell, {\it {Higher Spin Gravity Amplitudes From Zero-form
  Charges}},  \href{http://xxx.lanl.gov/abs/1208.3880}{{\tt arXiv:1208.3880}}.

\bibitem{Didenko:2012tv}
V.~Didenko and E.~Skvortsov, {\it {Exact higher-spin symmetry in CFT: all
  correlators in unbroken Vasiliev theory}},  {\em JHEP} {\bf 1304} (2013) 158,
  [\href{http://xxx.lanl.gov/abs/1210.7963}{{\tt arXiv:1210.7963}}].

\bibitem{Didenko:2013bj}
V.~Didenko, J.~Mei, and E.~Skvortsov, {\it {Exact higher-spin symmetry in CFT:
  free fermion correlators from Vasiliev Theory}},  {\em Phys. Rev.} {\bf D88}
  (2013) 046011, [\href{http://xxx.lanl.gov/abs/1301.4166}{{\tt
  arXiv:1301.4166}}].

\bibitem{Didenko:2012vh}
V.~Didenko and E.~Skvortsov, {\it {Towards higher-spin holography in ambient
  space of any dimension}},  {\em J. Phys.} {\bf A46} (2013) 214010,
  [\href{http://xxx.lanl.gov/abs/1207.6786}{{\tt arXiv:1207.6786}}].

\end{thebibliography}\endgroup

\end{document}